\documentclass{JINST}
\usepackage{amsmath}

\title{Geant4 simulation of the PSI LEM beam line: energy loss and muonium formation in
  thin foils and the impact of unmoderated muons on the $\mu$SR spectrometer}

\author{Kim Siang Khaw$^a$\thanks{Corresponding
author.}~\thanks{Present address: Department of Physics, University of Washington, Seattle, WA 98195, USA}~, Aldo Antognini$^a$, Paolo Crivelli$^a$, Klaus Kirch$^{a,b}$, Elvezio Morenzoni$^b$, Zaher Salman$^b$, Andreas Suter$^b$ and Thomas Prokscha$^b$\\
\llap{$^a$}Institute for Particle Physics,\\
  ETH Z{\"u}rich, Switzerland\\
\llap{$^b$}Paul Scherrer Institute,\\
  Villigen, Switzerland\\
  E-mail: \email{khaw84@uw.edu}}

\abstract{The PSI low-energy $\mu$SR spectrometer is an instrument
  dedicated to muon spin rotation and relaxation
  measurements. Knowledge of the muon beam parameters such as spatial,
  kinetic energy and arrival-time distributions at the sample position
  are important ingredients to analyze the $\mu$SR spectra. We present
  here the measured energy losses in the thin carbon foil of the muon
  start detector deduced from time-of-flight measurements.  Muonium
  formation in the thin carbon foil (10~nm thickness) of the muon
  start detector also affect the measurable decay asymmetry and
  therefore need to be accounted for. Muonium formation and energy
  losses in the start detector, whose relevance increase with
  decreasing muon implantation energy ($<10$~keV), have been
  implemented in Geant4 Monte Carlo simulation to reproduce the
  measured time-of-flight spectra. Simulated and measured time-of-flight
  and beam spot agrees only if a small
  fraction of so called ``unmoderated'' muons which contaminate the
  mono-energetic muon beam of the $\mu$SR spectrometer is introduced.  Moreover the
  sensitivity of the beam size and related upstream-downstream
  asymmetry for a specially shaped ``nose'' sample plate has been studied for various beam line settings,
  which is of relevance for the study of thermal muonium emission into vacuum from mesoporous
  silica at cryogenic temperatures. }


\begin{document}
\section{Introduction}

\begin{figure}[tbh]
\centering
\includegraphics[width=\textwidth]{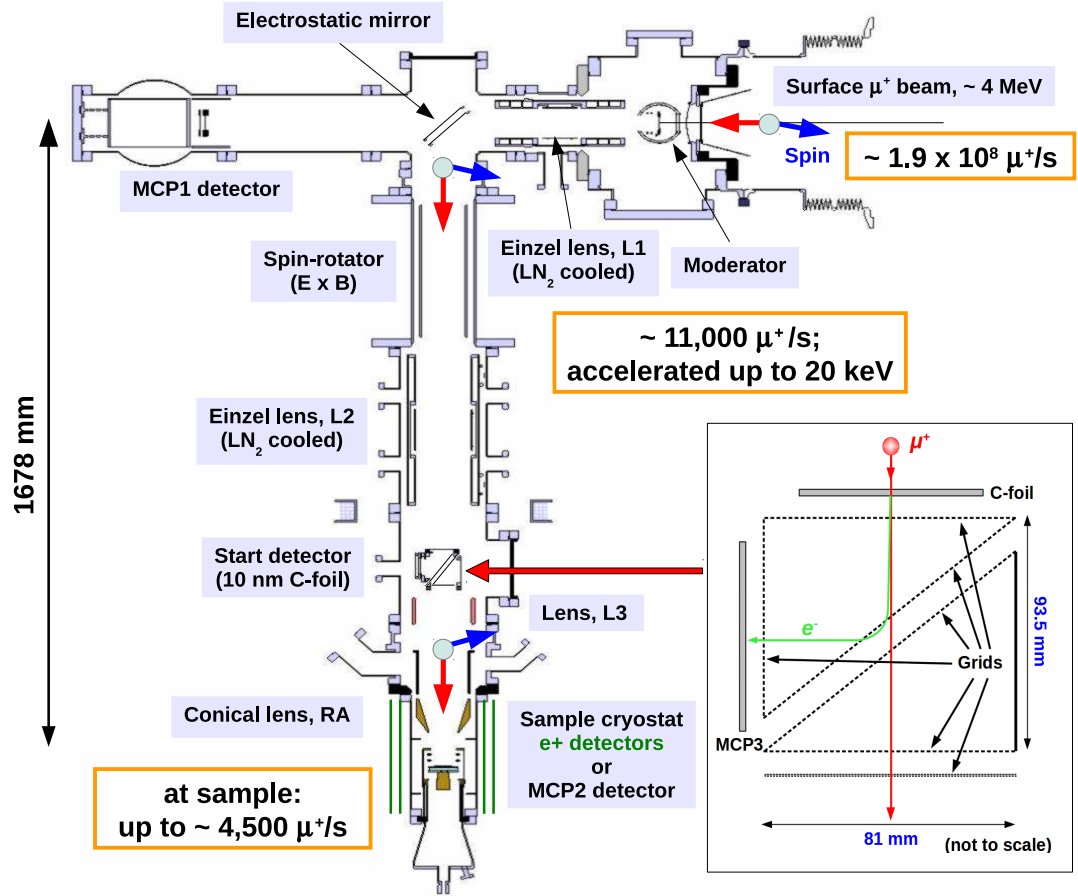} 
\caption{Schematic of the new LEM beam line~\cite{Lem2014}. The 4~MeV surface
  $\mu^{+}$ is moderated at an efficiency of about 0.01\% to an energy
  of about 15~eV before being re-accelerated again to energies up to
  20~keV. The $\mu^{+}$ beam is then bent by a 45$^\circ$
  electrostatic mirror before going through the spin rotator and the
  start detector and arriving at the sample plate mounted on the cold
  finger of the cryostat.}
\label{fig:lem2012}
\end{figure}

Polarized positive muons $\mu^{+}$ can be used to investigate structural
properties and dynamical processes of solid states via so called
$\mu$SR technique~\cite{Yaouanc2010} which stands for a collection of methods as Muon
Spin Rotation, Relaxation and Resonance.
A low-energy $\mu^{+}$ beam with tunable energy in the keV regime~\cite{Morenzoni1994}
plays an important role in the $\mu$SR field because these muons can be used to
investigate thin films.
The Low Energy Muon (LEM) beam line at PSI shown in
figure~\ref{fig:lem2012} is delivering $\mu^{+}$ with tunable kinetic
energy between 0.5 and 30 keV allowing the study of thin films and
multi-layers as a function of the implantation depth ranging from
0.5~nm to a few hundred~nm~\cite{Morenzoni2000,Morenzoni2004}.
The low-energy $\mu^{+}$ are obtained by moderating a surface $\mu^{+}$ beam
(4 MeV energy) from the $\mu$E4 beam line~\cite{Prokscha2008} with a 125~$\mu$m
thick Ag foil coated with a 200-300~nm thick layer of solid Ar-N$_{2}$~\cite{Prokscha2001}.
The moderated $\mu^{+}$ leaving the solid Ar-N$_{2}$ surface have mean kinetic
energy of 15~eV.
As the moderator is placed at high voltage (typically
$V_{\rm{mod}}=15$~kV), after leaving the moderator the $\mu^{+}$ are
accelerated to about 15~keV kinetic energy.
Using various electrostatic elements, the $\mu^{+}$ are transported
from the moderator to the sample region. Neglecting the various
focusing elements (einzel lenses and conical lenses), the $\mu^{+}$ are
first bent by a 45$^\circ$ electrostatic mirror and then transported through a
spin rotator. 
Subsequently, they cross a thin carbon foil (C-foil) which acts as a start detector 
before being implanted into the sample. The nominal density and thickness of the C-foil
are $\sim$2~$\mu$g/cm$^2$ and 10~nm, respectively.

While crossing the C-foil the $\mu^{+}$ is ejecting several electrons (on average 3) of few eV
energy from the foils surface whose detection define the $\mu^{+}$ implantation
time in the sample and the start of the event in the data acquisition.
The initially mono-energetic $\mu^{+}$ arriving at the start detector
undergoes energy and angular straggling which degrade the beam quality and
affect the measured $\mu$SR time spectra.
Moreover, by traversing the thin C-foil, a fraction of the $\mu^{+}$ can
undergo charge-exchange and leave the foil as muonium (Mu) or
negatively charged muonium (Mu$^{-}$) which decrease the measurable
total muon decay asymmetry.

Another complication is represented by the low-energy tails of the
$\mu^{+}$ leaving the Ar-N$_{2}$ moderator not as epithermal $\mu^{+}$ at the eV energy
but as only partially moderated muons whose energy is sufficiently low to be
deflected by the 45$^\circ$ electrostatic mirror and transported to
the sample region.
We term these muons as ``unmoderated'' muons. 

All these processes, at the moderator and at the C-foil, affect the
kinetic energy distribution of the $\mu^{+}$ leaving the C-foil and
consequently the $\mu^{+}$ arrival time distribution at the sample
position (relative to the signal in the start detector).
This arrival time distribution needs to be known to understand the
detailed shape of the decay positron time spectra at early times.

In this paper, measurements of the energy loss in the thin C-foil at various
$\mu^{+}$ energies which have been done using a
time-of-flight (TOF) technique are presented.
Energy losses and Mu formation in the 10~nm thick C-foil of the start
detector as well as ``unmoderated'' fraction of $\mu^{+}$ have been
implemented in the musrSim~\cite{Sedlak2012} simulation package which is
 based on Geant4~\cite{Agostinelli2003} to match the measured
TOF spectra.
The Geant4 simulation has been performed starting from the downstream of the
moderator till the sample region.

The LEM beam line was upgraded in 2012 to allow for longitudinal
$\mu$SR measurements, to have a better suppression of the proton/ion
background from the moderator and to have a better time resolution.
The first two items have been achieved by the installation of a spin
rotator~\cite{Salman2012} as shown in figure~\ref{fig:lem2012} after the
electrostatic mirror, whereas the improved time resolution has been
achieved by reducing the distance between the start detector and the
sample region (from 1164~mm to 563~mm distance, c.f. figure~1 of~\cite{Morenzoni2000}).

The beam sizes at the sample position for variations
of the beam line settings have been studied and compared with
determinations obtained from the upstream-downstream asymmetry
measurements.

\section{Energy loss in the thin C-foil via TOF measurements}\label{sec:five-four}

To determine the energy loss in the C-foil of the start detector a
TOF technique was applied.
A muon beam with well defined kinetic energy given by the moderator high voltage (HV) is
focused into the C-foil of the start detector which is set to a
negative HV ($V_{C}= -3.38$~kV).
The secondary electrons knocked out in the downstream direction by the muon
crossing the foil are first accelerated and then deflected by a system of grids
towards a micro-channel-plate (MCP3) as shown in the inset of figure~\ref{fig:lem2012}.
The MCP3 signal provides the ``start'' signal of the TOF
measurement.
Another micro-channel-plate (MCP2) is placed at the position usually
taken by the sample holder while performing the $\mu$SR measurements.
A signal from the MCP2 delivers the stop time of the TOF measurement.
In addition, the detection of an e$^+$ from muon decay in the positron counters of the
$\mu$SR spectrometer is required.
\begin{figure}[tbp]
\centering
\includegraphics[width=0.75\textwidth]{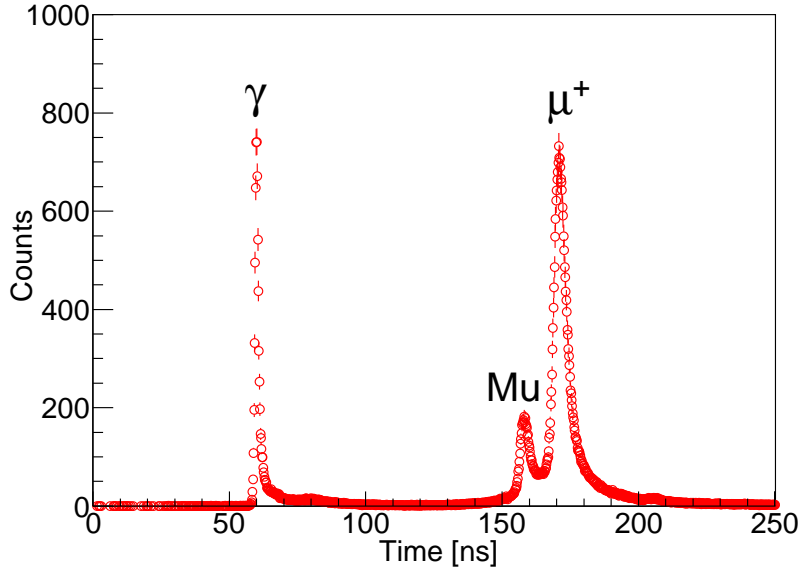}
\caption{TOF spectrum from the
  start detector to a micro-channel plate (MCP2) detector placed at the
  sample plate position for 12~keV $\mu^{+}$ beam transport energy. The
  peaks are corresponding to the prompt photons, the C-foil Mu atoms
  and the $\mu^{+}$, respectively. The time axis is set with an arbitrary zero.
  The HV at the RA and L3 lenses are switched off.}
\label{fig:TOF}
\end{figure}
The measured time spectra obtained in this way are shown in
figure~\ref{fig:TOF}.
Understanding these time spectra requires some knowledge of the
processes occurring in the C-foil.
The muons can leave the foil at various charge
states: $\mu^{+}$, Mu and Mu$^{-}$, as depicted in figure~\ref{fig:muonyield}(a).
The equilibrium yield for these various charge states scaled from
proton data~\cite{Gonin1994} are shown in figure~\ref{fig:muonyield}(b).
As the model of H formation (via overlap of the atomic states and solid electron states,
and electron tunneling given in \cite{Gonin1994}) when a low-energy
proton beam crosses a thin C-foil depends only on the velocity,
velocity scaling of the proton data has been assumed to calculate the Mu
charge state yield.
\begin{figure}[tbp]
\includegraphics[width=0.465\textwidth]{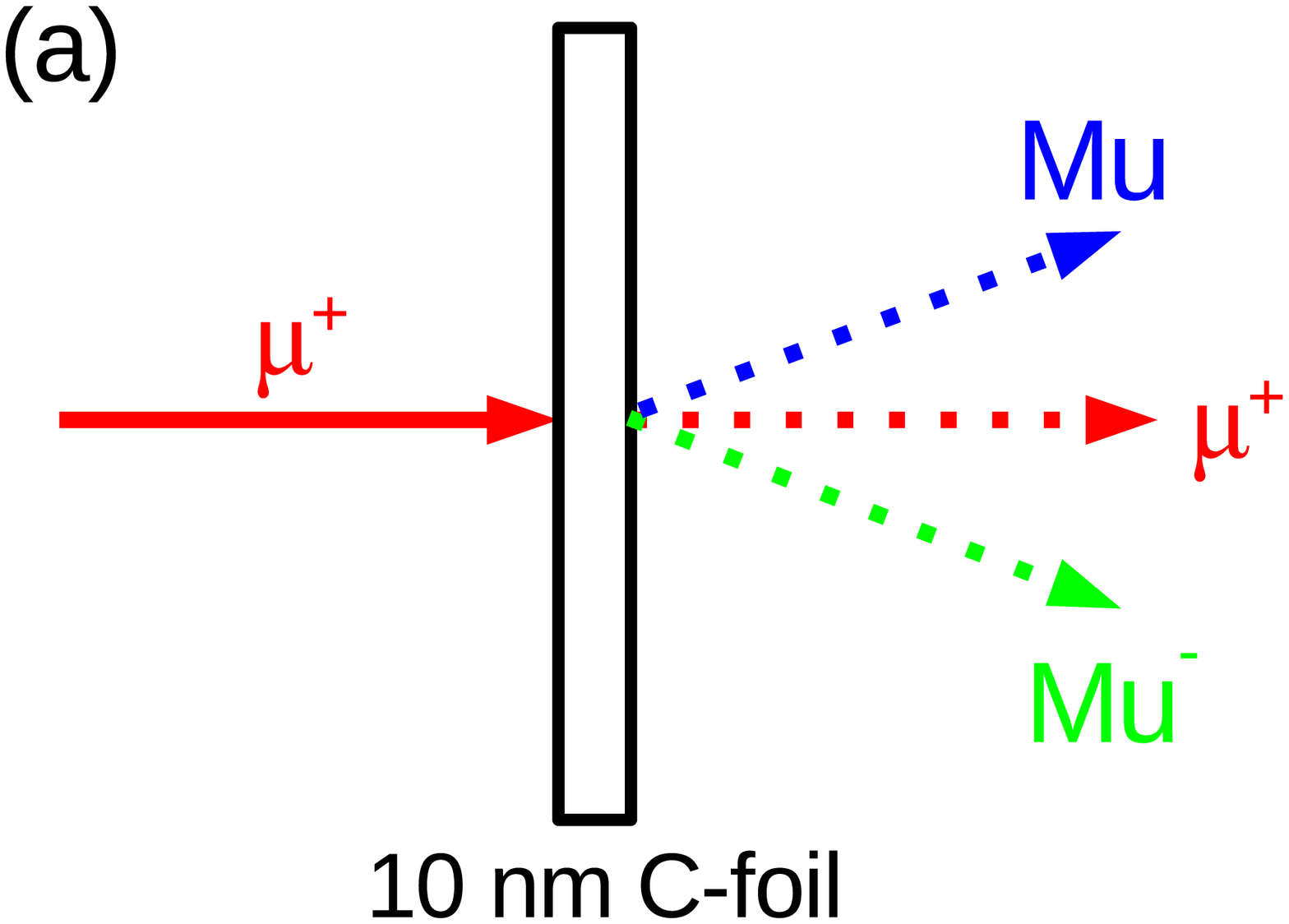}
\includegraphics[width=0.525\textwidth]{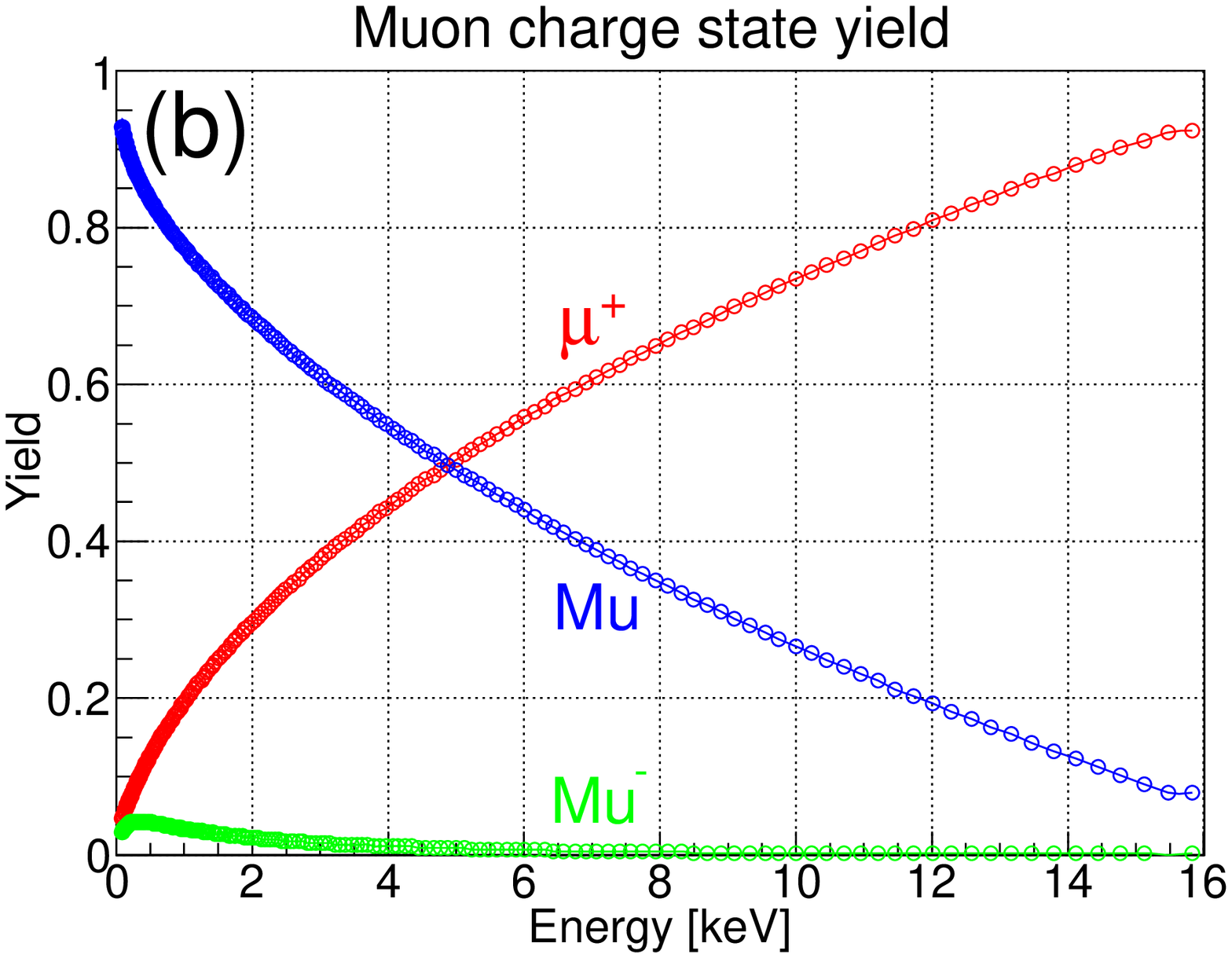}
\caption{(a) Formation of Mu and Mu$^{-}$ in the carbon foil due to
  the charge exchange process. (b) The charge state yields of the $\mu^{+}$ exiting the carbon foil 
  as a function of incoming muon energy at the foil surface, according to a velocity scaling of the
  proton data parameterizations~\cite{Gonin1994}.}
\label{fig:muonyield}
\end{figure}

The first peak at early times ($\gamma$ peak) is ensued in the
following way:
the electrons knocked out by a muon crossing the C-foil are
transported and detected in the MCP3 delivering the start time of the
event. The avalanche process occurring in the channel walls of the MCP3 generates about
10$^{6}$ electrons impinging with energies of a few hundred eV on the anode of MCP3,
where they generate UV photons by ionization/recombination processes or Bremsstrahlung.
Some of these photons could be detected (non-zero solid-angle acceptance and non-zero detection efficiency)
in the MCP2 providing the stop time of the event. The width of the $\gamma$ peak is about 0.7~ns and it corresponds
to the timing resolution of the TOF system (MCPs and electronics).
It was demonstrated that the position of this peak is not affected by
HV variations at the RA conical lens or at the L3 einzel lens or in
the start detector grids confirming that it is related with photon
emission in the MCP3.

The second peak is caused by Mu atoms traveling from the C-foil to
the MCP2 while the third (largest) peak is caused by $\mu^{+}$.
Also for these peaks the start signal of the events is given by the
electrons emitted from the C-foil.
Since the Mu motion is not affected by the electric fields in the
start detector, the Mu peak position (relative to the $\gamma$ peak)
can be related in a simple way to the kinetic energy of the Mu after
the C-foil.
The delayed timing of the $\mu^{+}$ peak compared to the Mu peak is
mainly due to the fact that $\mu^{+}$ has to overcome the negative
electrostatic potential in the start detector.
Thus, $\mu^{+}$ leave the start detector at a smaller kinetic energy
compared to Mu atoms producing the observed delay.

The time $t_{0}$ at which a $\mu^{+}$ is crossing the C-foil is given by:
\begin{equation}
t_{0} = t_{\gamma } - \Delta t_{\rm{FE}} -  \Delta t_{c}~,
\end{equation}
where $t_{\gamma}$ is the position of the $\gamma$ peak, 

$\Delta
t_{\rm{FE}}$ is the TOF of the knocked out electrons from the
C-foil to MCP3 and $ \Delta t_{c}=1.67$~ns is the TOF of a particle with the
speed of light from the MCP3 to the MCP2.
A $\Delta t_{\rm{FE}}=13.5(5)$~ns has been determined from the time
difference of two prompt peaks in the $\mu^{+}$ decay time spectra in regular
$\mu$SR measurements.
The earlier peak is caused by positrons hitting the C-foil and producing foil electrons
successively detected in the $e^{+}$ counters surrounding the sample region 
and the later peak is caused by positrons hitting the MCP3 directly
and then detected in the $e^{+}$ counters.
The TOF of foil electrons to the MCP3 is independent of the $e^{-}$
emission position at the C-foil but has a small dependence on the emission angle.
However, this effect and the possible variation of the $e^{-}$ emission energy are included
in the uncertainty of $\Delta t_{\rm{FE}}$. The TOF of $\mu^{+}$ through TD was simulated
for various initial position on the C-foil. The resulting TOF spread is about 0.1~ns.
Since the motion of the Mu atom, being a neutral system, is not
affected by the electric field in the start detector or other electric
fields downstream of the start detector, the Mu atom will move from
the C-foil to the MCP2 with uniform velocity ($v_{\rm{Mu}}$).
Its TOF from the C-foil to the MCP2 ($\Delta
t^{\rm{meas}}_{\rm{Mu}}$) determined from the TOF spectra is
given by
\begin{equation}
\Delta t^{\rm{meas}}_{\rm{Mu}}=t_{\rm{Mu}}-t_0~,
\end{equation} 
where $t_{\rm{Mu}}$ is the position (most probable value) of the Mu
peak.
Therefore, the Mu kinetic energy right after crossing the C-foil is
given by
\begin{equation}
E^{\rm{CFoil}}_{\rm{Mu}} = \frac{m_{\rm{Mu}}}{2}v_{\rm{Mu}}^2 =\frac{m_{\rm{Mu}}}{2}\left (\frac{L}{\Delta t^{\rm{meas}}_{\rm{Mu}}}\right )^{2}~,
\label{eq:kin-energy}
\end{equation} 
where $m_{\rm{Mu}}$ is the Mu mass and $L$ the distance between the
C-foil and MCP2.
It can be assumed that both the $\mu^{+}$ and the Mu leaving the
C-foil have the same kinetic energy $E_{\rm{CF}}=E^{\rm{CFoil}}_{\rm{Mu}}$.
With this assumption, the $\mu^{+}$ energy loss in the C-foil
(independent on the muon charge state when leaving the foil) is simply
given by
\begin{equation}
E_{\rm{loss}}=eV_{\rm{mod}}-eV_{C}-E_{\rm{CF}}~,
\label{eq:eloss}
\end{equation}
where $V_{\rm{mod}}$ is the HV at the moderator and
$V_{C}$ the HV at the C-foil. 

\begin{figure}[tbp]
\centering
\includegraphics[width=0.75\textwidth]{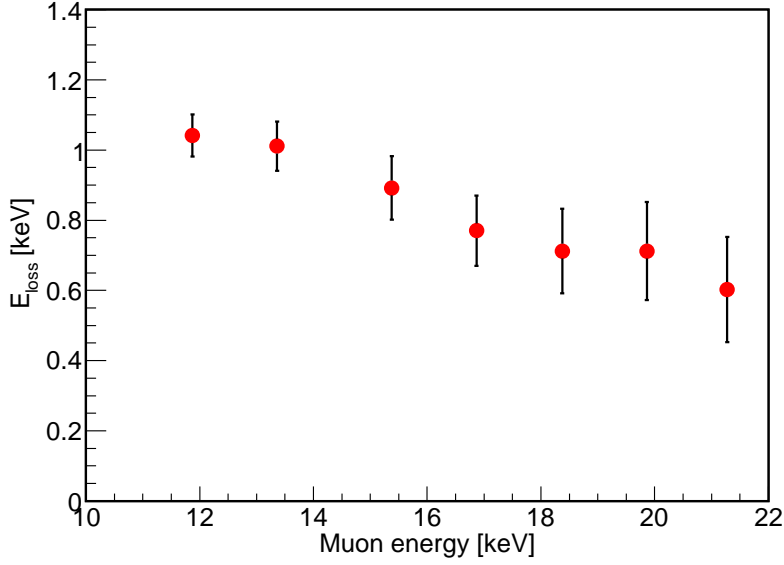}
\caption{Energy loss in the 10~nm C-foil determined from the Mu peak
  in the TOF spectra for various incoming muon energy at the foil surface.}
\label{fig:elosseimpl}
\end{figure}

The energy loss in the $\sim$2~$\mu$g/cm$^2$ C-foil for
various $\mu^{+}$ energies are summarized in table~\ref{ttof} and are
plotted in figure~\ref{fig:elosseimpl}. From the energy loss, a stopping power of $S=0.52(6)$ keVcm$^2$/$\mu$g
for a muon energy of 12~keV is obtained which has to be compared with the
value of 0.70(1)~keVcm$^2$/$\mu$g in~\cite{Hartmann1996}.
The uncertainty of the extracted stopping power and its deviation
from the value in~\cite{Hartmann1996} is related to the
uncertainty of the C-foil area density (0.2~$\mu$g/cm$^2$)
originating from the non-uniformity of the foil and the uncertainty
in the thickness from production to production.

\begin{figure}[tbp]
\centering
\includegraphics[width=0.9\textwidth]{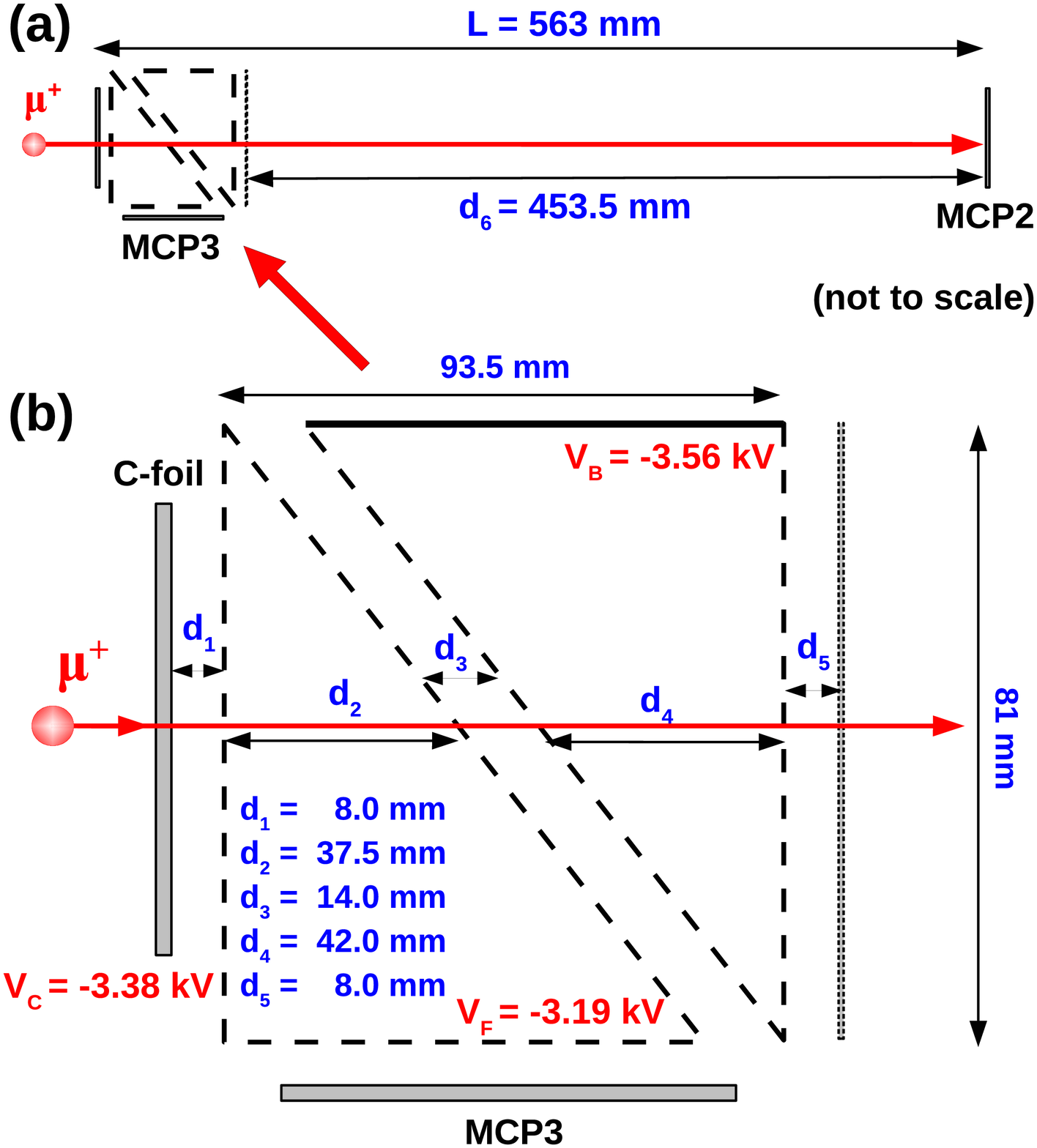}
\caption{(a) Schematic view of the TOF measurement. (b) Schematic of the start detector with various grids to deflect
the electrons and defined regions of constant potential or constant electric field. Distances $d_i$ ($i=1-6$) are defined for $\mu^{+}$
on the beam axis.}
\label{fig:startdetectorTOF}
\end{figure}

To check for the correctness of the assumed distance $L=563$~mm~ used
in Eq.~(\ref{eq:kin-energy}) we calculate the TOF of the $\mu^{+}$
from the C-foil to the MCP2 ($\Delta t^{\rm{calc}}_{\mu^{+}}$)
using simple kinematic calculations and compare it with
the TOF of $\mu^{+}$ peak determined directly from the measured TOF
spectra ($\Delta t^{\rm{meas}}_{\mu^{+}}$),
e.g. figure~\ref{fig:TOF}.
In order to calculate the $\mu^{+}$ TOF from the C-foil to the MCP2,
we need to consider the various regions defined by the distances $d_i$
as shown in figure~\ref{fig:startdetectorTOF}.
Since the TOF measurements have been accomplished without any HV at
the RA conical lens, L3 einzel lens and no electric fields in the
sample region the $\mu^{+}$ traveling from the C-foil till the MCP2
experience only regions of constant electric potential and two small
regions in the start detector of constant electric field.
The non-relativistic equations of motions for the various regions
$d_i$ are given by:
\begin{align}
d_{1} &=\frac{1}{2}a_{1}t^{2}_{1}+v_{1}t_{1}~, &v_{1}&=\sqrt{\frac{2E_{1}}{m}}~, &\Delta V_{1}&=(V_{C}-V_{F})~, &a_{1}&=\frac{q}{m}\frac{\Delta V_{1}}{d_{1}}~, \\
d_{2} &=v_{2}t_{2}~, &v_{2}&=\sqrt{\frac{2E_{2}}{m}}~, &E_{2}&=E_{1}+\Delta V_{1}, &a_{2}&=0~, \\ 
d_{3} &= \frac{1}{2}a_{3}t^{2}_{3}+v_{3}t_{3}~, &v_{3}&=v_{2},&\Delta V_{3} &= (V_{F}-V_{B})~, &a_{3}&=\frac{q}{m}\frac{\Delta V_{3}}{d_{3}}~, \\
d_{4} &=v_{4}t_{4}~, &v_{4}&=\sqrt{\frac{2E_{4}}{m}}~, &E_{4}&=E_{2}+\Delta V_{3}~, &a_{4}&=0~, \\ 
d_{5} &= \frac{1}{2}a_{5}t^{2}_{5}+v_{5}t_{5}~, &v_{5}&=v_{4}~, &\Delta V_{5} &= (V_{B}-0)~, &a_{5}&=\frac{q}{m}\frac{\Delta V_{5}}{d_{5}}, \\
d_{6} &=v_{6}t_{6}~, &v_{6}&=\sqrt{\frac{2E_{6}}{m}}~, &E_{6}&=E_{4}+\Delta V_{5}~, &a_{6}&=0~ 
\end{align}
where $m$ is the $\mu^{+}$ mass, $E_{i}$ the $\mu^{+}$ kinetic energy at
the entrance of the $i$-region ($E_{1}=E_{\rm{CF}}$), $q$ the
charge of the particle, and $V_{C}=-3.38$~kV, $V_{F}=-3.19$~kV and
$V_{B}=-3.56$~kV the various HVs applied at the grids of the start detector.
The TOF $t_i$ in these various $i$-regions can be calculated using
these simple relations:
\begin{eqnarray}
t_{i} &=& \frac{d_{i}}{v_{i}} \text{~~~~~~~~for $a_{i}=0$}  \\
t_{i} &=& -\frac{v_{i}}{a_{i}} + \text{sign}(a_{i})\sqrt{\left(\frac{v_{i}}{a_{i}}\right)^{2}+\frac{2d_{i}}{a_{i}}}
\end{eqnarray}
where $v_{i}$ is the $\mu^{+}$ velocity when entering the region $d_i$,
and $a_{i}$ the $\mu^{+}$ acceleration in the region $d_i$ of constant
electric field.
The total TOF $\Delta t^{\rm{calc}}_{\mu^{+}}$ is eventually given by the sum
\begin{equation}
\Delta t^{\rm{calc}}_{\mu^{+}} = \sum^{6}_{i=1}t_{i}~.
\end{equation}

\begin{table}[tbp]
\caption{Moderator potential $V_{\rm{mod}}$, incoming muon energy at the C-foil surface $E_{\mu^{+}}$, measured and calculated TOF (for the peak maximum) of
  $\mu^{+}$ and Mu from the start detector to MCP2. The energy
  of the particle after the carbon foil $E_{\rm{CF}}$ is
  determined from $\Delta t^{\rm{meas}}_{\rm{Mu}}$ . Knowing the $E_{\rm{CF}}$, the energy loss
  $E_{\rm{loss}}$ in the carbon foil can be calculated using Eq.~(\protect\ref{eq:eloss}).
  The $\Delta t^{\rm{calc}}_{\mu^{+}}$ assumes the distances $d_1$ to $d_6$ given by mechanical construction and also
  assumes mono-energetic $\mu^{+}$ hitting the C-foil with energy given by the moderator and C-foil electric potentials, $E_{\mu^{+}}=eV_{\rm{mod}}+3.38$~keV.}
\centering
\begin{tabular}{|c|c|c|c|c|c|c|c|}  \hline
 $V_{\rm{mod}}$  (kV)            &  8.5     & 10.0 & 12.0 & 13.5 & 15.0 & 16.5 & 18.0 \\ \hline
 $E_{\mu^{+}}$ (keV) & 11.88 & 13.38 & 15.38 & 16.88 & 18.38 & 19.88 & 21.38 \\ \hline
 $E_{\rm{CF}}$ (keV)               & 10.84(6) & 12.37(7) & 14.49(9) & 16.11(10) & 17.67(12) & 19.17(14) & 20.78(15) \\ \hline
 $E_{\rm{loss}}$ (keV)                & 1.04(6)  & 1.01(7) & 0.89(9) & 0.77(10) & 0.71(12) & 0.71(14) & 0.60(15) \\ \hline
 $\Delta t^{\rm{meas}}_{\rm{Mu}}$ (ns)       & 131.4(5) & 123.0(5) & 113.7(5) & 107.8(5) & 102.9(5) & 98.8(5) & 94.9(5) \\ \hline
 $\Delta t^{\rm{meas}}_{\mu^{+}}$ (ns)  & 152.7(5) & 139.8(5) & 126.4(5) & 118.6(5) & 112.1(5) & 106.8(5) & 101.6(5) \\ \hline
 $\Delta t^{\rm{calc}}_{\mu^{+}}$ (ns)  & 152.9(3) & 140.0(3) & 126.5(3) & 118.4(3) & 112.0(3) & 106.7(3) & 101.8(3) \\ \hline
\end{tabular}
\label{ttof}
\end{table}

As can be seen by comparing the 5-th with the 6-th rows of table~\ref{ttof}, there is a very good agreement between $\Delta
t^{\rm{calc}}_{\mu^{+}}$ and $\Delta t^{\rm{meas}}_{\mu^{+}}$, confirming the consistency of our TOF analysis, including the
correctness of all distances and the assumption that Mu and $\mu^{+}$ exiting the C-foil have same kinetic energies $E_{\rm{CF}}$.

\section{Geant4 simulation of the energy losses in a thin carbon foil}

\begin{figure}[htbpp]
\centering
\includegraphics[width=\textwidth]{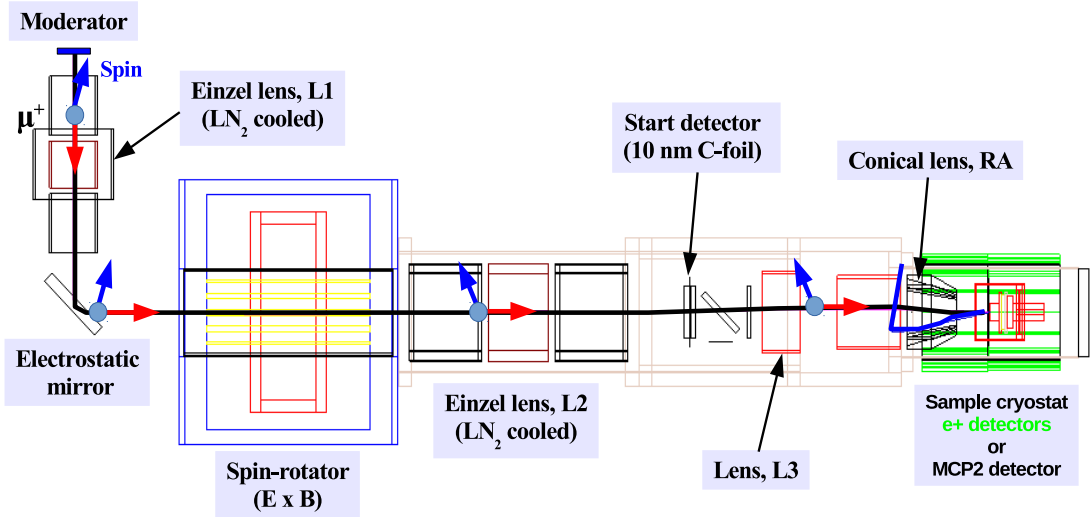} 
\caption{Schematic view of the LEM beam line as implemented in the Geant4 simulation. The beam which is given by the black solid line is simulated starting after the moderator. The arrows indicate spin (blue) and momentum (red) directions.}
\label{fig:LEM_withSR_sim}
\end{figure}

A Geant4-based (version 9.4 patch 04) simulation of the beam
propagation in the LEM beam line has been accomplished.
Figure \ref{fig:LEM_withSR_sim} shows the geometry implemented in the Geant4 simulation.
Beam line components relevant to this study are included, from the L1 einzel lens,
spin rotator (SR) to the sample chamber. The detailed geometry including radiation shield,
sample holder and cold finger of the cryostat are also implemented.

Precise electric and magnetic field maps are inputs to the Geant4 Monte Carlo simulation.
The magnetic field map of the SR was measured in a volume $5\times5\times5$~cm$^{3}$ around the origin of the SR coodinate system.
The electric field of the SR was calculated using the commercial OPERA
finite element programs (TOSCA/OPERA-3D)~\cite{Opera2014}, and the electric field maps
were calculated with the finite element software COMSOL~\cite{Comsol2014}.
Due to the modified sample plate setup in this work described in Sec. 4, the electric field maps
of the conical lens (RA) and the copper sample plate with cylindrical nose were re-calculated.
The electrostatic module of COMSOL was used and a fine mesh was applied for higher accuracy calculations.
A 2~mm spacing of grid points are used for the electric field maps.
An example of the electric potential map in the sample region is shown in figure~\ref{fig:Efieldcomsol}.
Initial conditions of the $\mu^{+}$ beam is summarized in table~\ref{tab1}. The typical number of events
generated is ranging from $10^{6}$ to $10^{7}$, such that the statistical uncertainty is reduced to less than 1\%.

The low-energy physics processes currently not available in
Geant4 have been implemented to describe the energy losses and Mu
formation processes in the thin C-foil of the start detector.
When a $\mu^{+}$ is impinging on the C-foil our Geant4 simulation
performs following operations:
\begin{itemize}
\item Decide about the charge state of the exiting muon, between
  $\mu^{+}$, Mu and Mu$^{-}$ using the yields given in
  figure~\ref{fig:muonyield}.
\item Calculate the energy loss (same for all particle charge state)
      assuming a Landau distribution  with most probable value (MPV)
      given by the energy loss determined from the TOF measurements. 
\item Compute angular scattering using the standard Geant4 package for
  multiple-scattering.
\end{itemize}
More details regarding these three operations are given in the following subsections.

\subsection*{Formation of muonium in the carbon foil}
Formation of ``foil'' Mu at the thin C-foil is implemented by velocity
scaling of existing data from proton--C-foil
data~\cite{Gonin1994,Hofer1998,Paraiso2006} as shown in
figure~\ref{fig:muonyield}.
These ``foil'' Mu will be stopped when they reach a material
interface.

\subsection*{Muon energy loss in the carbon foil}

In Geant4.9.4, models simulating the $\mu^{+}$ energy loss
and its fluctuation are implemented in the C++ class
{\it G4MuIonisation}. 
By default, for $\mu^{+}$ energy below 200~keV,
{\it G4BraggModel} is used where the energy losses are derived from the
tabulated stopping power for proton using velocity scaling~\cite{Allisy1993}.
Energy loss fluctuation of $\mu^{+}$ is
simulated by means of the {\it G4IonFluctuation} model. 
For a thin absorber, the energy fluctuation is based on a very simple two energy-level
atom model and the particle-atom interaction give rise either to an atomic
excitation or an atomic ionization with energy loss distributed according
to~\cite{Bichsel1988}.

In our simulation, the $\mu^{+}$ energy loss is simulated based on the
values determined from the TOF measurements and its fluctuation is
implemented using a Landau random number generator based on
CERNLIB~\cite{Kolbig1984}. 
It was found by L. Landau that a certain linear function of the energy loss has, under certain
assumptions, a universal (i.e. parameter free) density~\cite{Landau1944}. 
The generated random number from the universal Landau distribution $X$
is first shifted to have only positive energy losses ($X+3.5$) and
then scaled linearly so that its MPV coincides with the measured $E_{\rm{loss}}$
in table~\ref{ttof}, i.e. the randomly generated energy loss distribution is
\begin{equation}
E^{\rm{random}}_{\rm{loss}} = \Big(\frac{X+3.5}{3.5}\Big)\cdot E_{\rm{loss}}~.
\label{eq:Landau}
\end{equation}
Here, it is thus assumed that the energy loss distribution goes down to zero.

\begin{figure}[tbp]
\centering
\includegraphics[width=0.75\textwidth]{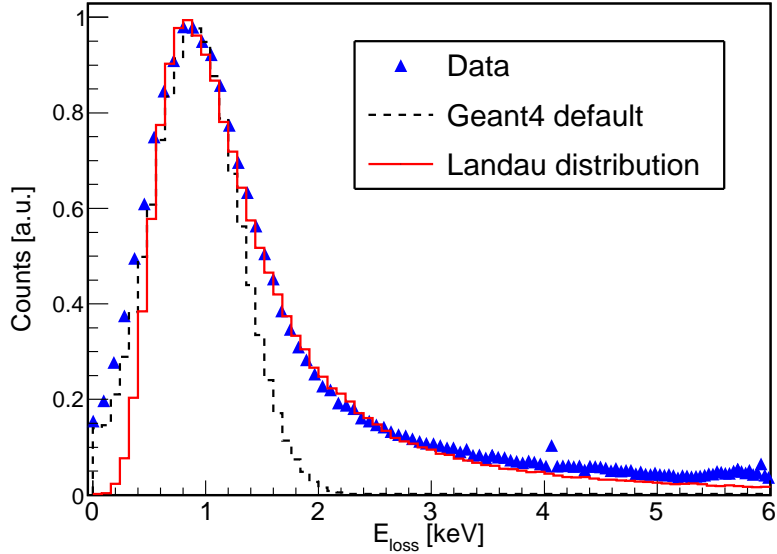}
\caption{Energy loss distributions of 15.38~keV muons in a 10~nm thick C-Foil with a density $\sim$2~$\mu$g/cm$^2$, obtained using the default Geant4
  package (dotted black line) or our extension based on Landau distributed energy losses (red solid line)
  where the MPV energy loss ($E_{\rm{loss}}=0.89$) is taken from the TOF measurements. The energy loss spectrum
  extracted from the TOF spectrum is also shown (blue triangles).
}
\label{fig:ElossCompare}
\end{figure}

A comparison between energy losses extracted from the TOF spectrum
and simulated energy losses using standard or our extended Geant4
version are shown in figure~\ref{fig:ElossCompare}.
A better agreement between simulations and measurements is
achieved when parameterizing the energy losses using
the Landau distribution of Eq.~(\ref{eq:Landau}).
Interestingly, the Geant4 default simulation gives a better agreement below 0.5~keV.
A cutoff approach was tried, however no improvement was achieved for the fitting of the TOF spectra.
It is important to stress that the main attention is on the high-losses tail because it impacts the first few 100 ns of muSR measurements.

\subsection*{Muon multiple scattering in the carbon foil}

In previous versions of Geant4, the measured transmissions of $\mu^{+}$ beam
from the start detector till the sample were poorly reproduced
due to the underestimation of the multiple-scattering process from
C-foil~\cite{Paraiso2006}.
However, recent versions of Geant4 have better physics models of
multiple-scattering which reproduce correctly the Meyer
scattering~\cite{Hofer1998,Meyer1971}.
In this paper, multiple Coulomb scattering
is simulated by using {\it G4MuMultipleScattering} based on the model
from Urban~\cite{Ivanchenko2010}.

\subsection*{Validation of the simulations}

\begin{figure}[tbp]
\centering
\includegraphics[width=0.75\textwidth]{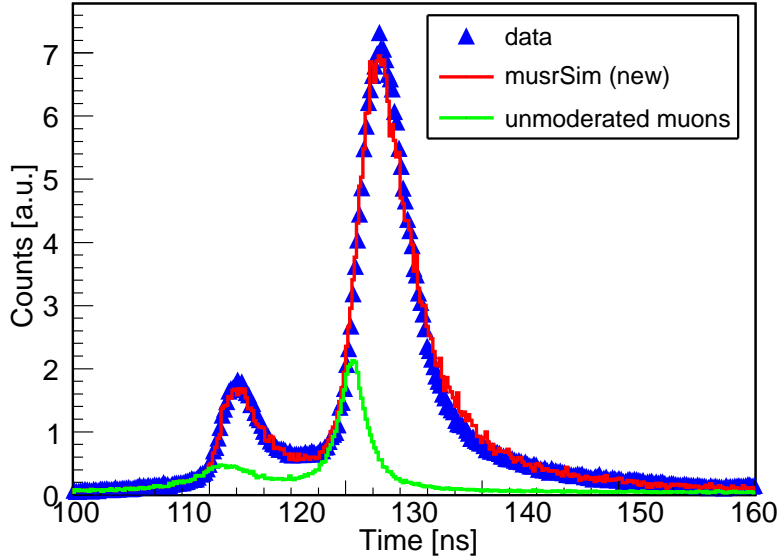}
\caption{Measured and simulated TOF. The green curve represents the
  contribution arising from ``unmoderated'' muons which has been added
  to the simulated TOF as an empirical sum of two ``Lorentzian'' to
  achieve good matching with the data (see text for details).}
\label{fig:TOFnew}
\end{figure}

To validate the implementation of the low-energy
processes, the simulated TOF spectra are compared to
the measured TOF spectra.
In figure~\ref{fig:TOFnew} a comparison between simulated and measured
TOF is presented.
The green curve corresponds to the
contribution of these ``unmoderated'' muons which have to be
assumed in order to match the measurements with the simulations.
These are muons leaving the moderator not as epithermal muons
at eV energy but as the non-fully moderated tail of the muon beam with keV energies.

About 40\% of the $\mu$E4 beam hits the moderator target where about one half
is stopped~\cite{Prokscha2008,Prokscha2001}. This means that about 20\% of the incoming $\mu^{+}$ beam will go through
the moderator as ``unmoderated'' $\mu^{+}$ with a mean energy of several hundred keV and
a low-energy tail ranging down to few keV energies~\cite{Prokscha1998}.
Even though a large fraction of the ``unmoderated'' $\mu^{+}$ will not
be reflected by the electrostatic mirror, they still contribute about 10-15\%,
depending on the moderator HV, to the total $\mu^{+}$ which are impinging on the sample.
It has been shown~\cite{Shiroka2000} that the electrostatic mirror which is set
at the same HV as the moderator ($V_{\rm{mod}}$) deflects particles
with kinetic energies ($E_k$) in the range $eV_{\rm{mod}}<E_k<2eV_{\rm{mod}}$ by 90$^\circ$.

Hence, ``unmoderated'' muons with slightly larger kinetic
energy compared with the ``moderated'' muons are deflected towards the
sample region.
This explains why the ``unmoderated'' fraction contribute to the time
spectrum at slightly earlier times compared with the ``moderated''
muons as well visible in Fig~\ref{fig:TOFnew}.

\begin{figure}[tbp]
\centering
\includegraphics[width=0.495\textwidth]{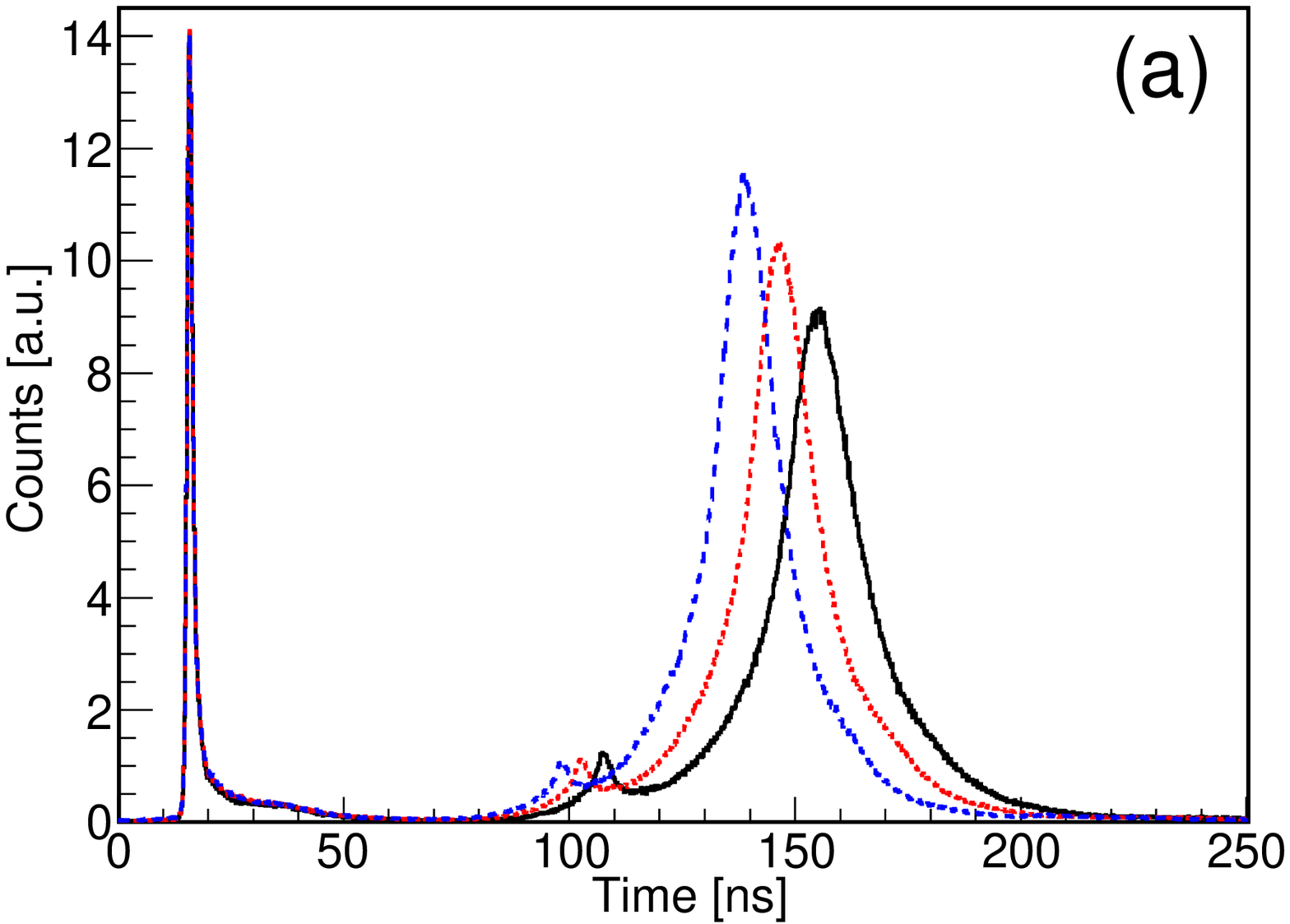}
\includegraphics[width=0.495\textwidth]{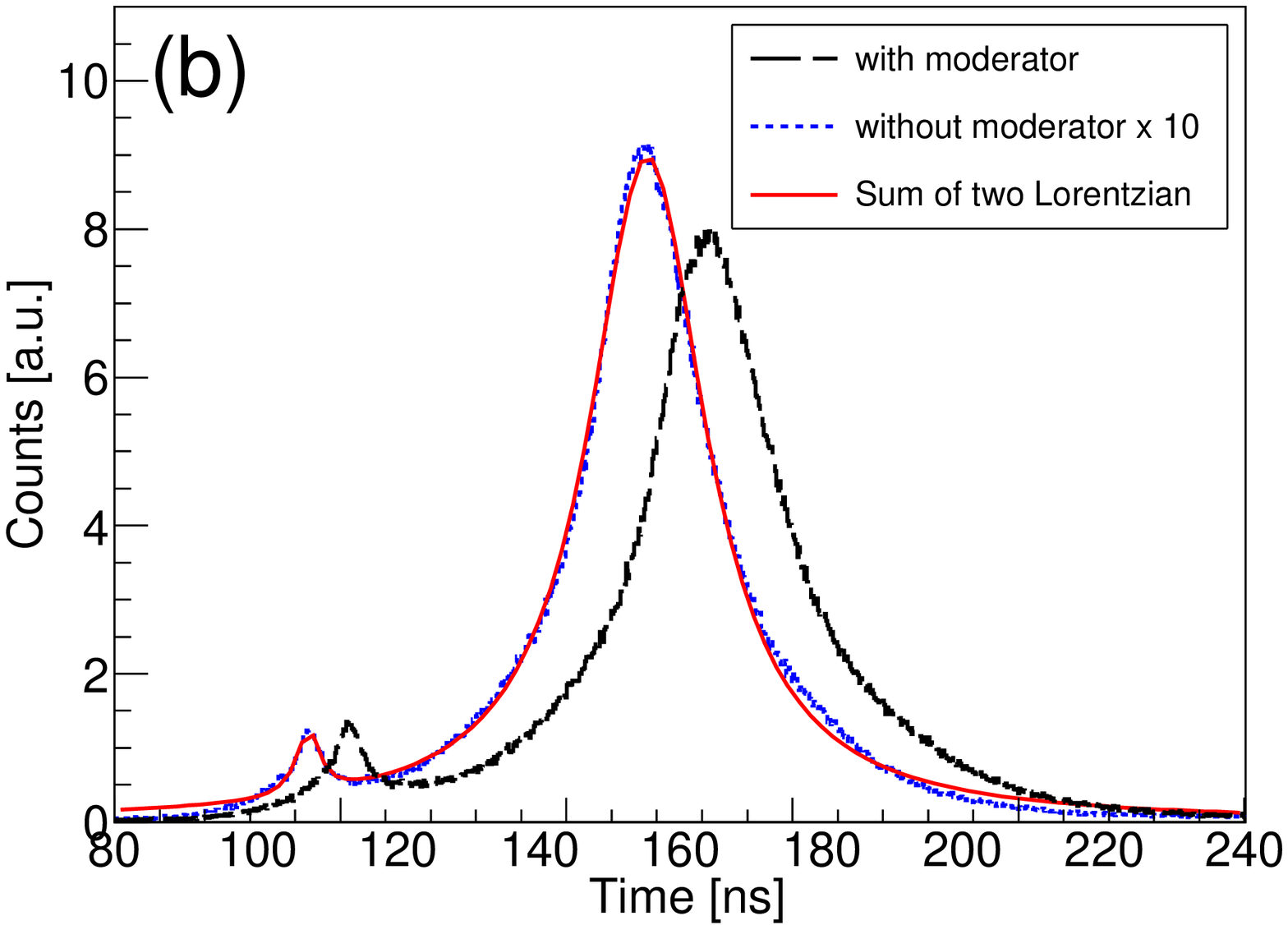}
\caption{(a) TOF spectra of the ``unmoderated'' $\mu^{+}$ determined
  from a measurement without any Ar-N$_{2}$ layer on the moderator for
  various moderator and electrostatic mirror HVs (solid black line $-12$ kV, dotted red line $-13.5$ kV and dashed blue line $-15$ kV). (b) Measured
  ``unmoderated'' TOF at 12 kV beam transport (blue dotted line) fitted with the sum of two Lorentzian
  functions (red solid line). For comparison a TOF spectrum (black dashed line) of
  ``moderated'' muons measured with the Ar-N$_{2}$ moderator rescaled by a
  factor of 10 is also shown. }
\label{fig:TOFnew-unmoderated}
\end{figure}

The existence of ``unmoderated'' muons has been experimentally
verified by taking data without any Ar-N$_{2}$ layer at the moderator.
The corresponding TOF spectrum is shown in
figure~\ref{fig:TOFnew-unmoderated}.
Differently from the situation in figure~\ref{fig:TOFnew} this
measurement was performed with non-zero voltage at the conical lens.
The contribution of ``unmoderated'' muons to the TOF spectra can be
empirically described by the sum of two Lorentzian functions which
account for the Mu and the $\mu^{+}$ peaks.
The ``unmoderated'' muons TOF were poorly reproduced by 
Geant4 simulation even after implementing the Landau energy loss 
distribution described in Sec. 3. This is due to the insufficient knowledge
of the phase space of these muons after exiting the moderator. An extensive
study is required to reproduce the shape of the TOF.

The ``unmoderated'' muon contribution given by the green curve in
figure~\ref{fig:TOFnew} results from the sum of two ``Lorentzian''.
The relative widths and amplitudes of this two ``Lorentzian'' peaks can
not be assumed from the ``unmoderated'' measurement because of the
different HV settings of the conical lens focusing the beam on the
MCP2 and therefore are free parameters.
In conclusion, simulated and measured TOF spectra agree very well
together if a Landau distributed energy straggling in the C-foil is
used and a small fraction of ``unmoderated'' muons is accounted for.
It is important to note that the TOF spectra of figure~\ref{fig:TOFnew}
at times around $t \in [108;112]$~ns and $t \in [118;122]$~ns cannot
be reproduced simply by modifying the muon energy losses in the C-foil assumed
in the Geant4 simulation. The data in these two regions can be reproduced only by the contamination
of ``unmoderated'' muons.

Simulations can then be used to determine the $\mu^{+}$ and Mu kinetic
energy distributions and related arrival time distributions at the
sample plate for any beam line settings
(moderator HV, sample HV, conical lens HV etc).
Slow $\mu^{+}$ and Mu tails cause detrimental distortions of the
measured $\mu$SR time spectra which need to be accounted for when
considering the ``early'' part of the measured time
spectra.
The starting point of the
time window where $\mu$SR fit can be reliably applied without being
distorted is dictated by the low-energy $\mu^{+}$ tail.

The knowledge of the fraction of ``foil'' Mu entering the LEM
spectrometer is also an essential input for the analysis of the
$\mu$SR data.
As the precession frequency of the muon spin in the Mu atoms is a
factor of 100 larger than that for a free muon~\cite{Hughes1970}, the ``foil'' Mu
produced at the C-foil give rise to depolarization effects and
reduction of the observable total decay asymmetry.
This is because the Mu eventually stop not only in the sample region
with a well defined constant magnetic field but also on the
thermal shield of the sample cryostat and other elements which may have depolarizing effects and
are placed at various B-field values. In the simulation,
we assume that once Mu has formed, depolarization occurs independently of its states (singlet or triplet).

The recent upgrade of the LEM beam line was characterized mainly 
by the insertion of the spin rotator and the
moving of the trigger detector closer to the sample region.
The closer placement of the trigger detector to the sample region
has caused a larger fraction of
the ``foil'' Mu to reach the sample region, enabling a
better study of the process related with ``foil'' Mu production.
The insertion of the spin rotator opened the way for longitudinal $\mu$SR
measurements, broadening the spectrum of possibilities available at the PSI-LEM
spectrometer.
In addition, the spin rotator was designed also to reduce 
beam contamination into the sample region.
However the insertion of the spin rotator changed in an still not
fully understandable way the beam propagation, degrading the beam size
at the sample position.
It is probably the insufficient knowledge of the fringe fields of the newly inserted spin
rotator which do not allow an exact simulation of the transport of the muons.

\section{Upstream-downstream asymmetries and the muon beam sizes}

\begin{figure}[tbp]
\includegraphics[width=0.36\textwidth]{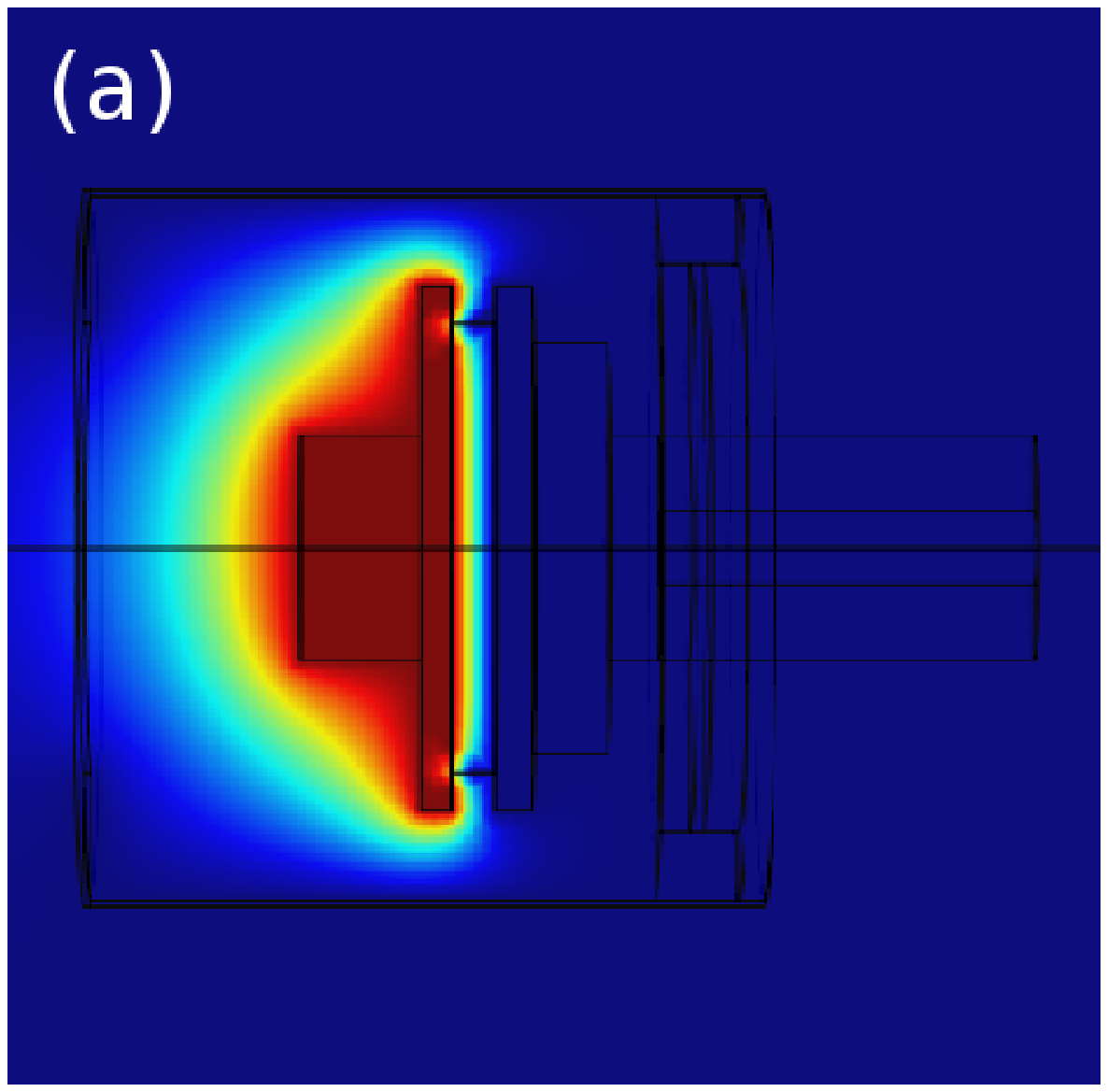}
\hfill
\includegraphics[width=0.62\textwidth]{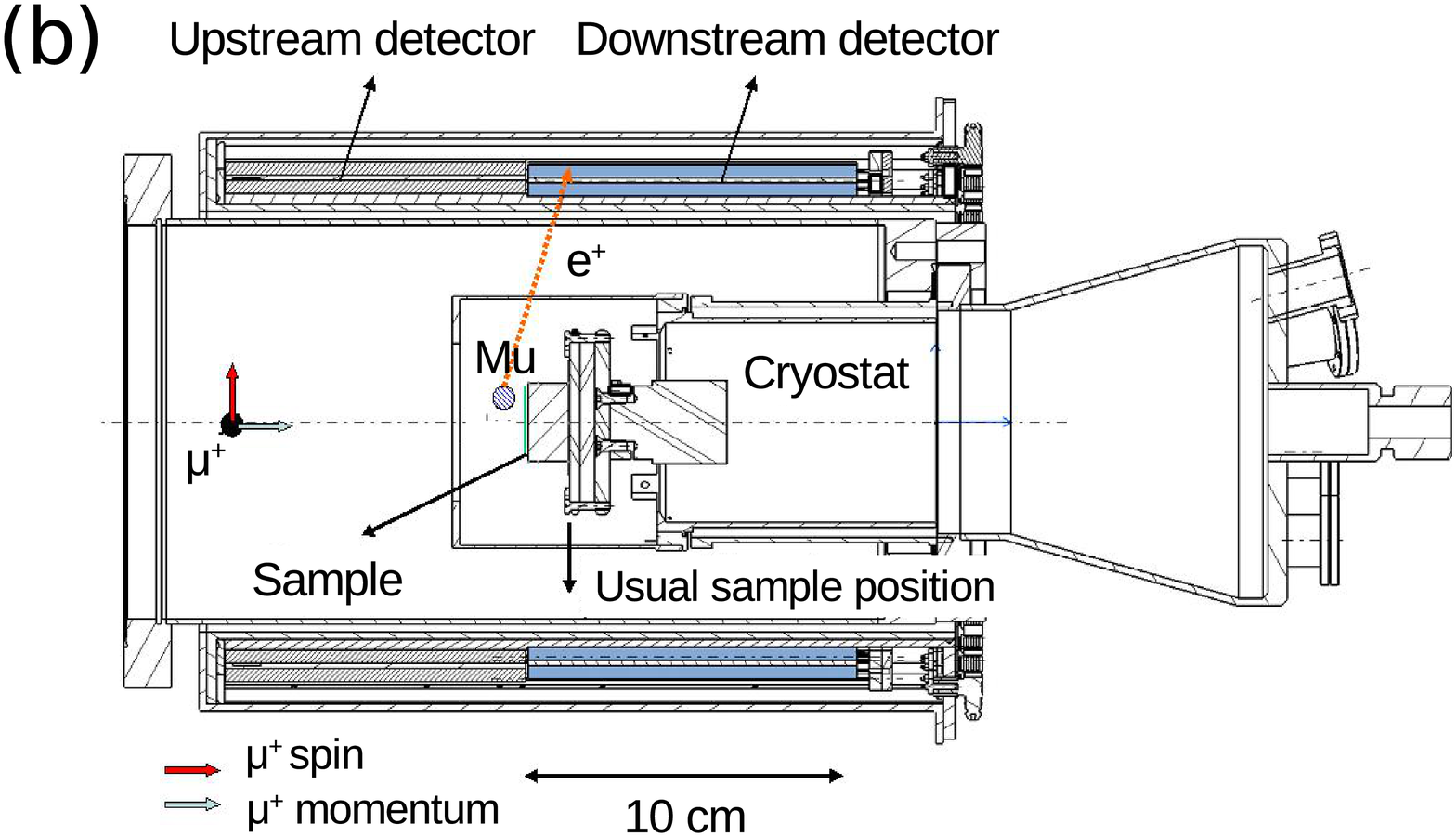}
\caption{(a) Electric potential contour plot in the ``nose sample plate'' region
  calculated with COMSOL multiphysics~\cite{Comsol2014}. The ``nose sample plate''
  was designed for the dedicated experiment of ~\cite{Antognini2012}. The normal
  sample plate for $\mu$SR experiments is a flat disk. (b) The sample plate region
  is surrounded with upstream and downstream positron counters.
  }
\label{fig:Efieldcomsol}
\end{figure}

\begin{figure}[tbp]
  \includegraphics[width=0.575\textwidth]{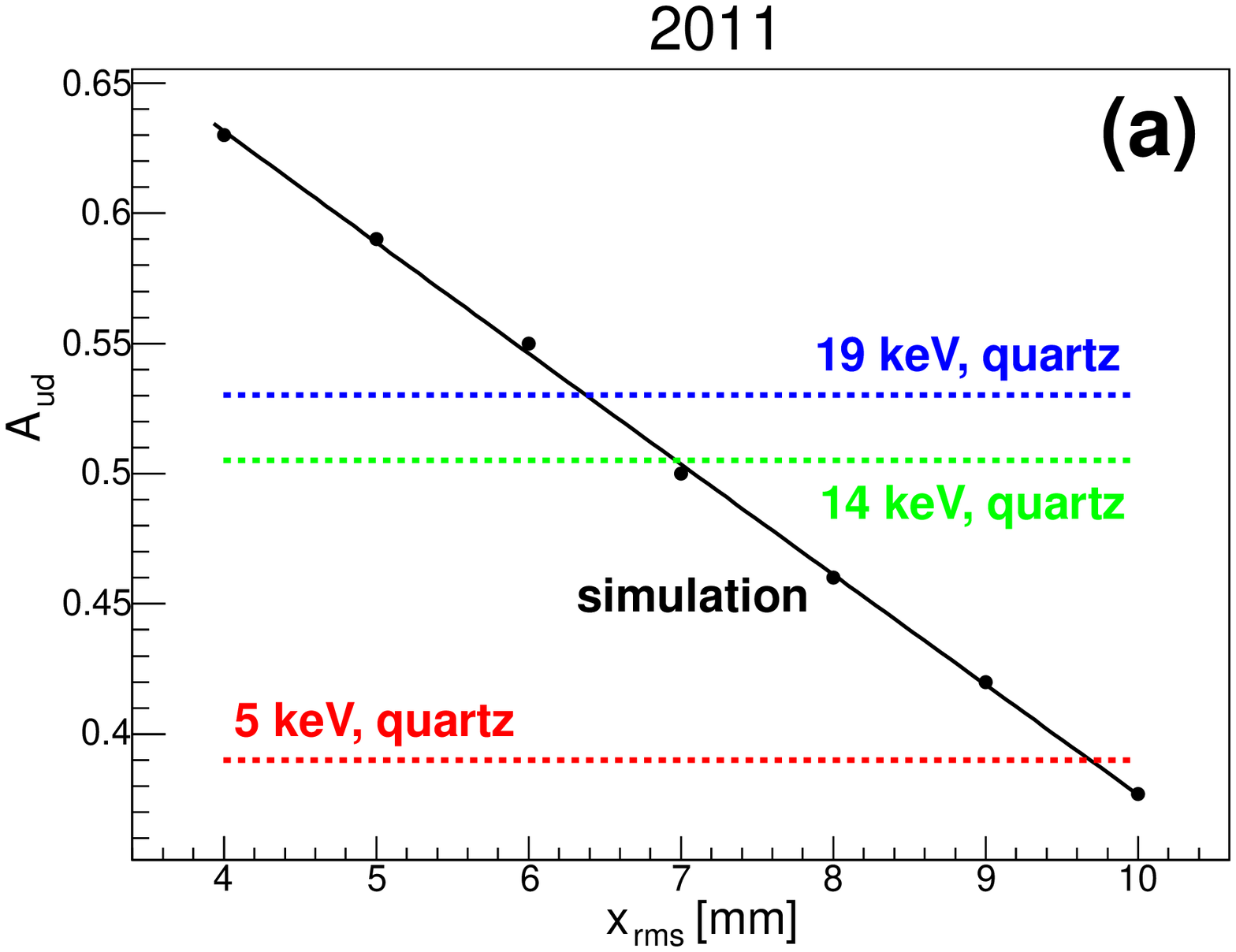}
  \includegraphics[width=0.42\textwidth]{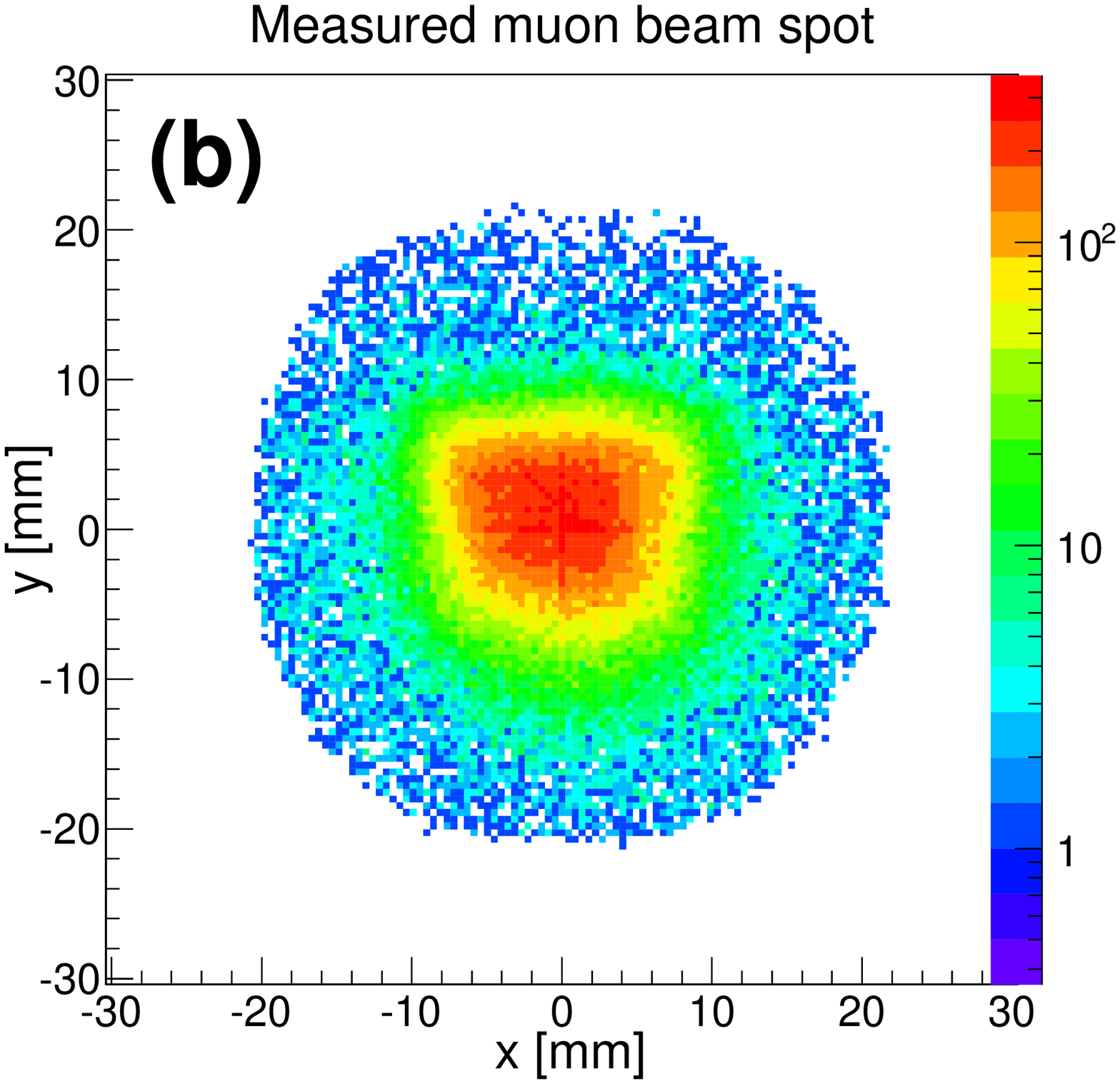}
  \includegraphics[width=0.575\textwidth]{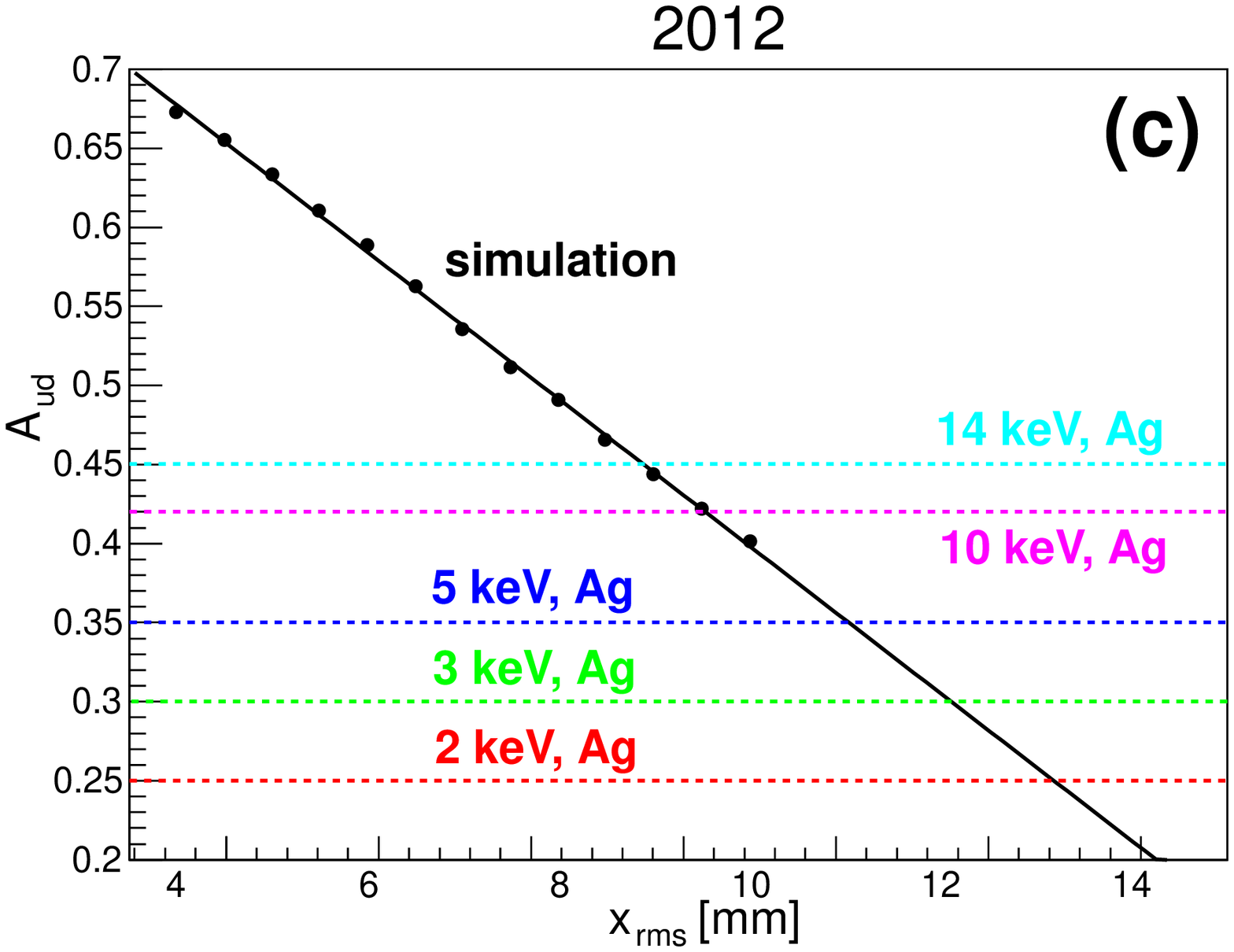}
  \includegraphics[width=0.42\textwidth]{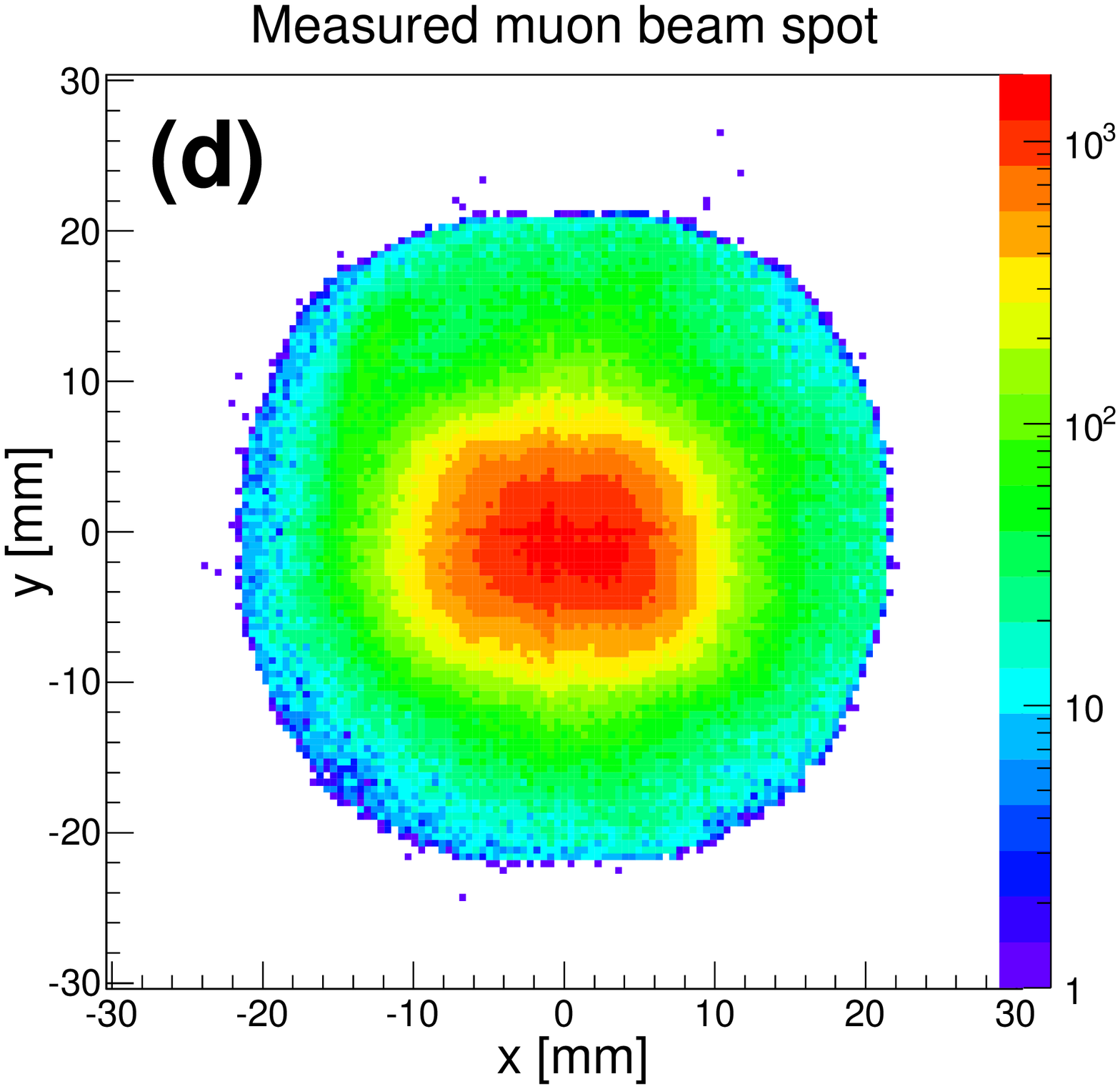}
  \caption{(a+c) The black dots are the simulated upstream-downstream
    decay asymmetry $A_{\rm{ud}}$ for various RMS values
    $x_{rms}$ of the muon beam spot. These plots are used to
    determine the experimental muon beam sizes at the sample
    position. In the simulation, Mu is generated at rest on top of (a) a
    1~mm thick $20 \times 20$~mm$^{2}$ fused quartz disk on top of the nose sample plate, for
    year 2011 setup, without spin rotator and (c) the
    Ag coated nose sample plate, year 2012 setup, with spin rotator
    in the low-energy muon beam line. The horizontal dotted lines are
    the measured decay asymmetries. (b+d) Muon
    beam spot at 14~keV implantation energy measured with the
    MCP2. The top (bottom) panels are before (after) the LEM
    upgrade. Quartz and Ag samples were used for the determination of
    muon beam sizes because Mu emission into vacuum is absent in these materials
    and $A_{\rm{ud}}$ are time independent.}
\label{fig:Audvssigma2011}
\end{figure}

The beam size at the sample position is needed to normalize and
analyze the $\mu$SR data but also to understand the LEM beam line and
validate the Geant4 transport simulation.
For example knowledge of the beam size at the sample position is used to remove
the contributions arising from muons not impinging on the sample of
interest.
The so-called upstream-downstream asymmetry $A_{\rm{ud}}$ can
be used to infer the beam spot size at the sample position:
\begin{equation}
A_{\rm{ud}}(t)=\dfrac{N_{\rm{u}}(t)-N_{\rm{d}}(t)}{N_{\rm{u}}(t)+N_{\rm{d}}(t)}~,
\label{eq:aud}
\end{equation}
where $N_{\rm{u,d}}(t)$ are the total number of decay positron detected as a function of time,
in the upstream and downstream detectors surrounding the sample region as shown
in figure~\ref{fig:Efieldcomsol}(b). The values of $A_{\rm{ud}}$ given in this paper were obtained by
fitting $A_{\rm{ud}}(t)$ with a constant function for times larger than 200~ns.
The fitted $A_{\rm{ud}}$ for various implantation energies are shown in figure~\ref{fig:Audvssigma2011}(a) and (c).
The upstream-downstream asymmetry and its time evolution is also the
central ingredient of the longitudinal $\mu$SR technique.

A MCP plate at the sample position can be used to perform not only
measurements of the muon TOF, but also measurements of the beam
profiles which are shown in figure~\ref{fig:Audvssigma2011}(b) and (d).
The MCP measurements provide a 2-dimensional profile of the muon beam,
but it can only be used when there is no HV applied to the sample plate.
On the contrary, the upstream-downstream asymmetry is 
strongly correlated with the beam size, and it can be
measured on-line and for any sample-plate HV.
This plays a crucial role especially at low energy, when a high
positive HV has to be applied to the sample plate ($V_{\rm{sample}}$) to tune the
$\mu^{+}$ implantation energy $E_{\rm{implant}}$ which is given by
\begin{equation}
E_{\rm{implant}} = eV_{\rm{mod}} - eV_{\rm{sample}} - E_{\rm{loss}}~.
\end{equation}
The electric potential ensued by the HV at the conductive sample plate of 
figure~\ref{fig:Efieldcomsol}(a) shows a curvature of the equipotential lines
which gives rise to a radial force causing a defocussing of the muon beam
and thus an increase of the muon profile at the sample plate.
Note that this effect is particularly relevant for the ``nose sample plate''
shown in figure~\ref{fig:Efieldcomsol} which was developed for a
dedicated experiment, which looked for thermal Mu emission into vacuum from mesoporous
silica targets~\cite{Antognini2012}.
The standard $\mu$SR sample holder, being a simple plate, do not show
such a strong curvature and therefore the beam defocussing effect is smaller.

A large variation of the beam size at the sample position for small
variation of the beam parameters when using this ``nose sample plate''
has been observed.
Because of this sensitivity a study of the $A_{\rm{ud}}$
asymmetry using this ``nose'' sample plate for various beam
line settings was performed to investigate the validity of the
Geant4 beam transport simulation, which is also relevant for the analysis of the experiment for thermal Mu
emission into vacuum~\cite{Antognini2012}.

The black lines in figure~\ref{fig:Audvssigma2011}(a) and (c) shows
the correlation between the beam size and the asymmetry $A_{\rm{ud}}$.
They have been computed assuming muon decaying from the sample plate
with a given transverse spatial distribution described by a 2D Gaussian
function with a width $x_{RMS}=y_{RMS}=\sigma_{x,y}$. Note that
$A_{\rm{ud}}$ can be larger than the theoretical-maximum-decay asymmetry
of 0.33 due to the shielding effect of the nose sample plate on the
downstream detector, i.e. positron has a lower probability of reaching the downstream detector
(according to Eq.~(\ref{eq:aud}), $A_{\rm{ud}} \rightarrow 1$ when $N_{\rm{u}} \gg N_{\rm{d}}$).

By comparing the measured asymmetries (horizontal dotted lines) with
the asymmetry versus beam size predicted from the simulations, the beam
size can be extracted.
On his turn this beam size can be compared with the beam size obtained
from a transport simulation of the full LEM beam line starting from the
moderator till the sample region, including the processes in the
C-foil and the electric fields in the modified sample region shown in
figure~\ref{fig:Efieldcomsol}(a).
The asymmetry measurement can be thus used to validate the Geant4 beam
transport of the LEM beam line.

Figure~\ref{fig:Audvssigma2011}(a) and (c) show a decrease of the
asymmetry for decreasing implantation energy revealing that the
beam size increases considerably with decreasing energy.
This has to be related to a substantial defocussing effect when the HV
at the sample is increased due to stronger electric fields and slower
muon velocity.

From the correlation line deduced from Geant4 simulation as depicted in
figure~\ref{fig:Audvssigma2011}(a), the muon beam size at 14~keV
implantation energy has a RMS value $\sigma_{x,y}=6.9$~mm.
This value compares well with the value measured with the MCP2 of 6.3~mm
(The MCP2 is placed 18.5~mm downstream of the nose sample plate 
and hence a slightly smaller beam spot is expected) and with the
value of 7.2~mm computed with Geant4 beam transport.

However for the 2012 measurements, after the LEM upgrade, the beam
size values obtained from the MCP measurements and the values from the
asymmetry measurement do not agree with the beam size obtained from
the Geant4 simulation of the beam propagation in the LEM beam line.
This is calling for a verification and detailed investigation of the
LEM beam line simulations.
To study the sensitivity of $A_{\rm{ud}}$ to various beam line
settings the sample holder of figure~\ref{fig:Efieldcomsol} is used.
The beam line settings of table~\ref{tab1} are for the SR setup, experimentally optimized.
\begin{table}[tbp]
\centering
\caption{LEM beam line initial settings for the Geant4
  simulations and initial beam parameters.}
\begin{tabular}{|l|c|c|}\hline
                & \multicolumn{2}{c|}{High voltage setting (kV)} \\\hline
                Moderator       & 12                     & 15                   \\\hline
Einzel lens, L1 & 7.19                   & 8.99                 \\\hline
Mirror          & 12                     & 15                   \\\hline
Spin rotator, $E^{SR}_{x}$          & 2.09                  & 2.29                \\\hline
Spin rotator, $B^{SR}_{z}$          & 62.3 G                 & 69.7 G               \\\hline
Einzel lens, L2 & 8.38                   & 10.484               \\\hline
Lens, L3        & 8.99                   & 11.483               \\\hline
Conical lens, RA  & 9.15                   & 11.9                 \\\hline
 & \multicolumn{2}{c|}{Initial beam parameters} \\\hline
Beam size, $\sigma_{x,y}$ & \multicolumn{2}{c|}{7.5~mm} \\\hline
Beam divergence, $\sigma_{x',y'}$ & \multicolumn{2}{c|}{2.0$^\circ$} \\\hline
\end{tabular}
\label{tab1}
\end{table}

\begin{figure}[tbp]
\centering
\includegraphics[width=0.495\textwidth]{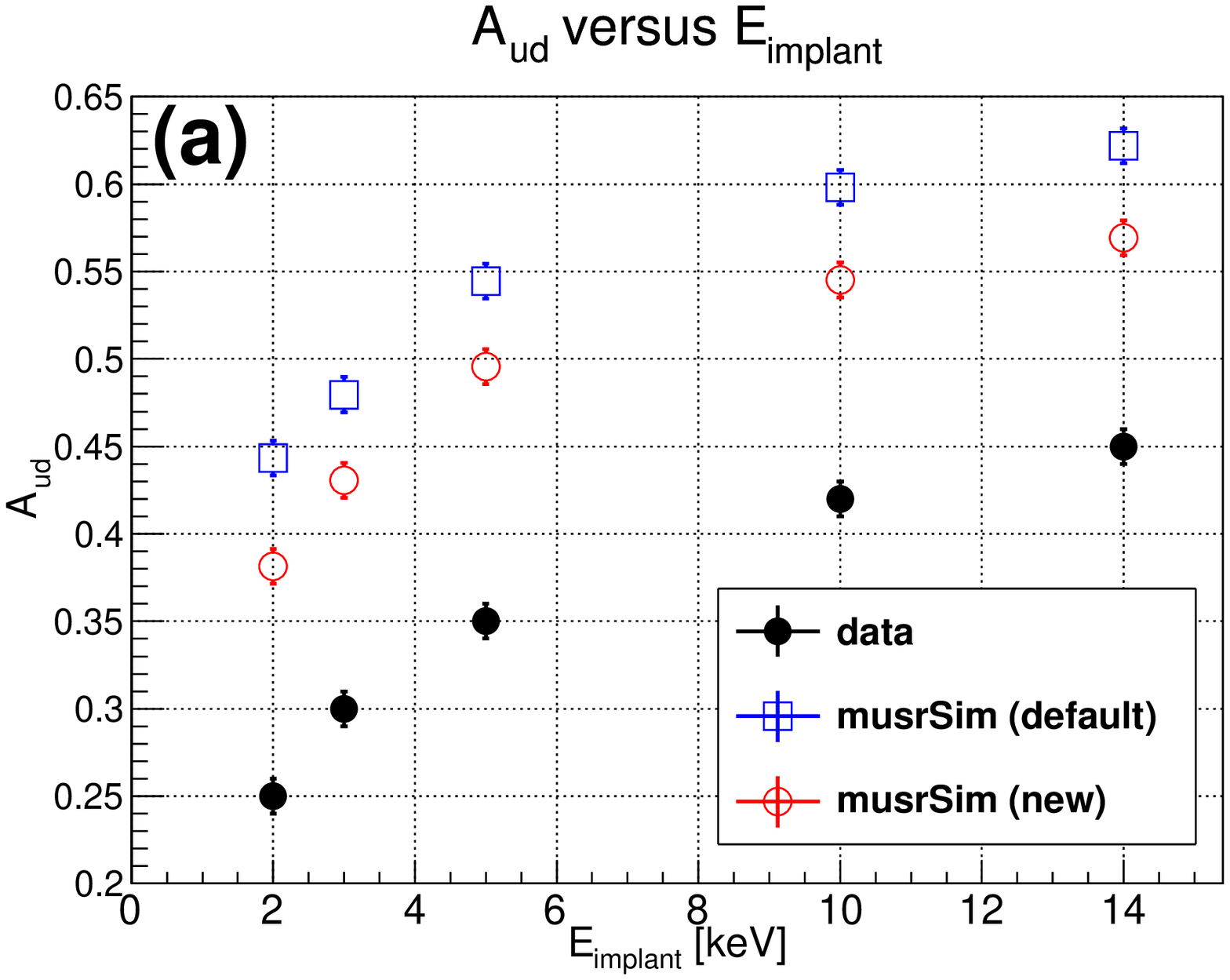}
\includegraphics[width=0.495\textwidth]{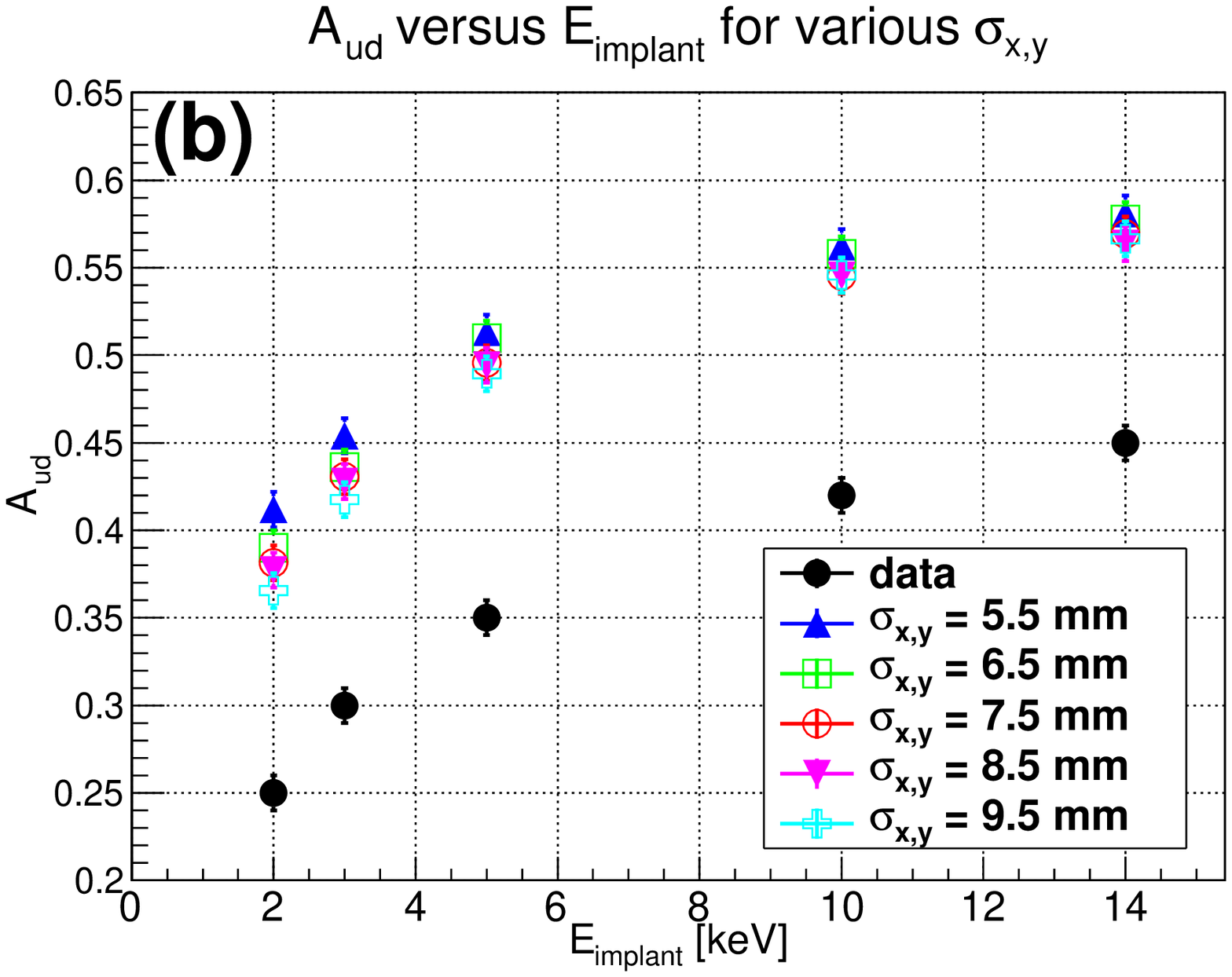}
\includegraphics[width=0.495\textwidth]{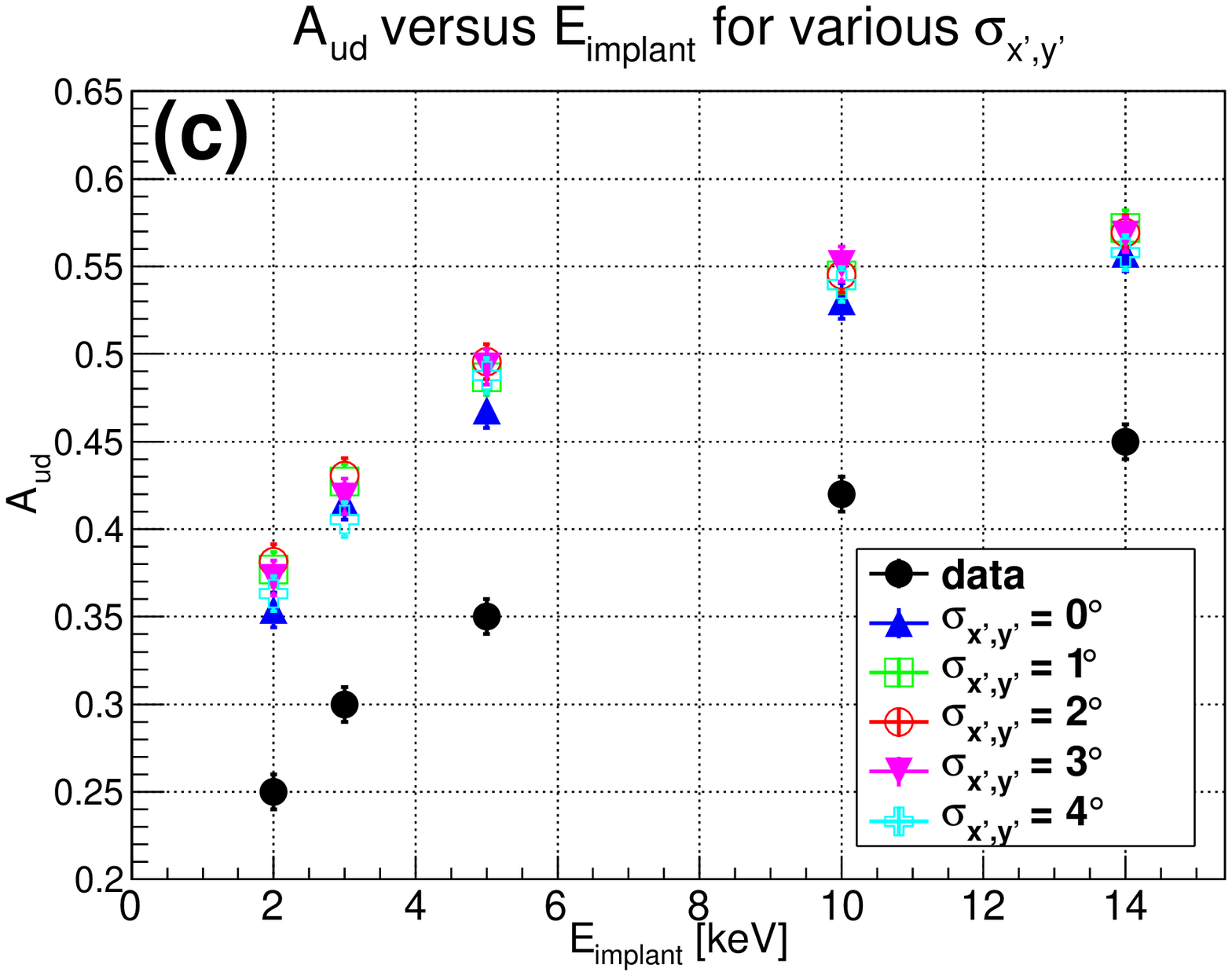}
\includegraphics[width=0.495\textwidth]{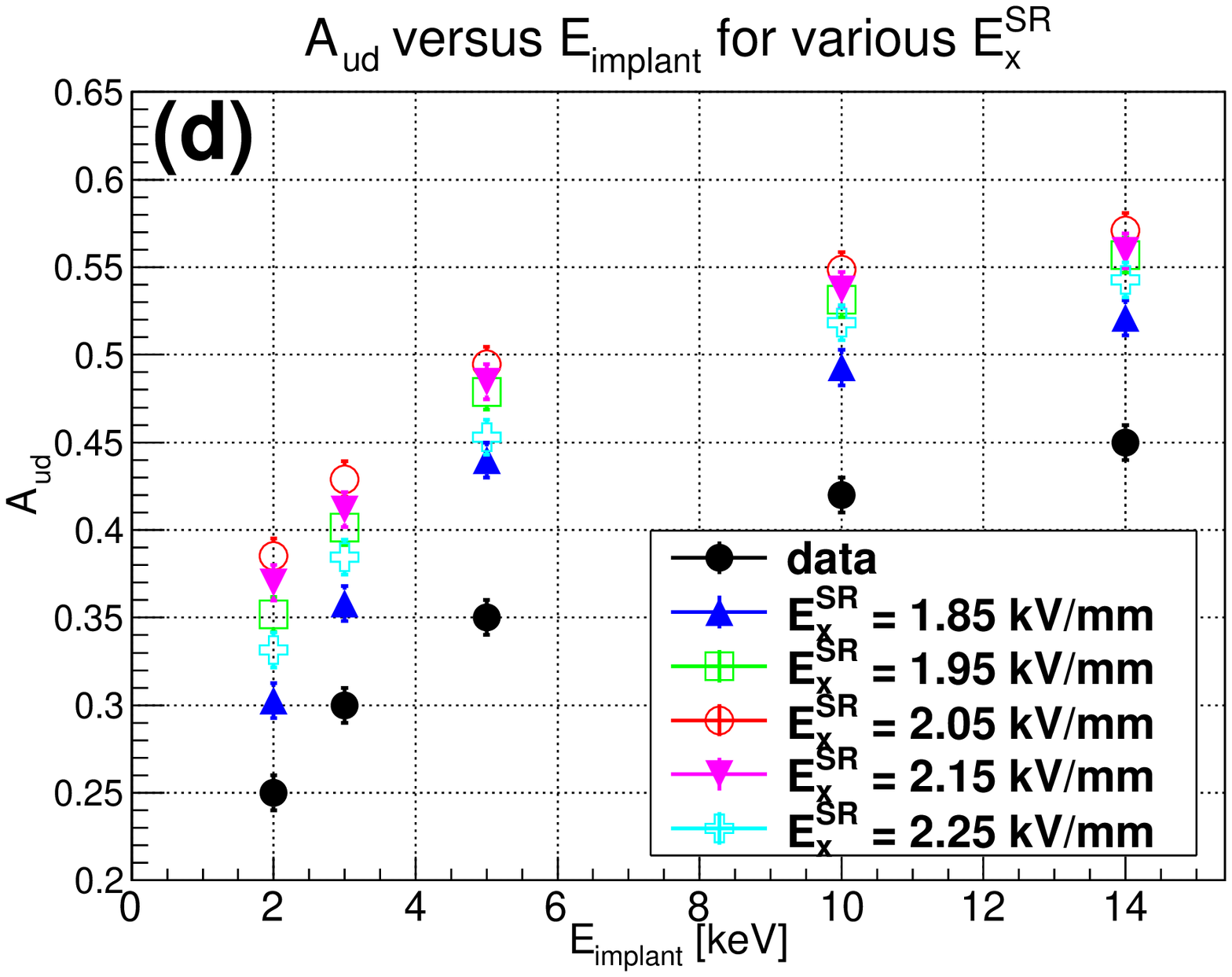}
\includegraphics[width=0.495\textwidth]{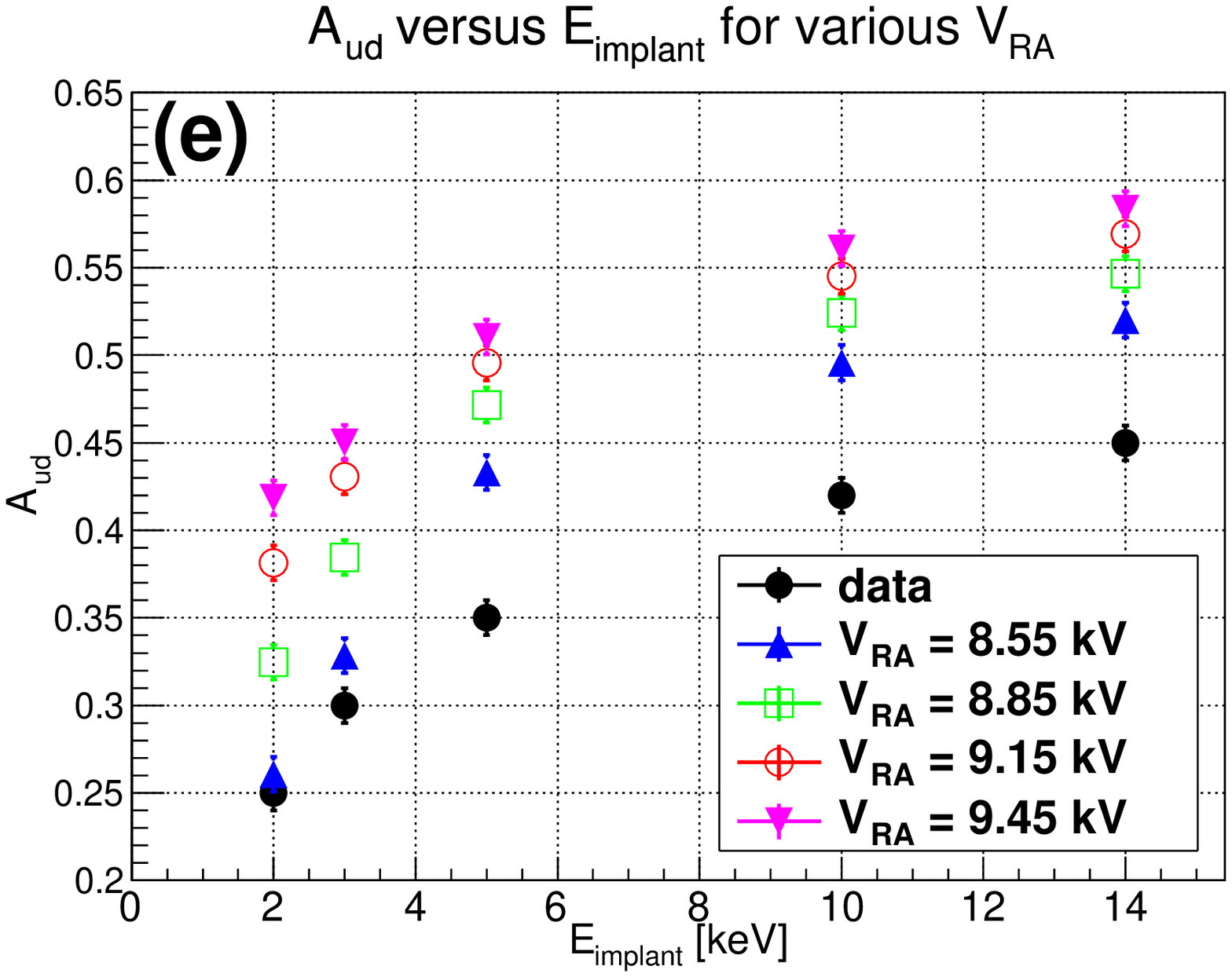}
\includegraphics[width=0.495\textwidth]{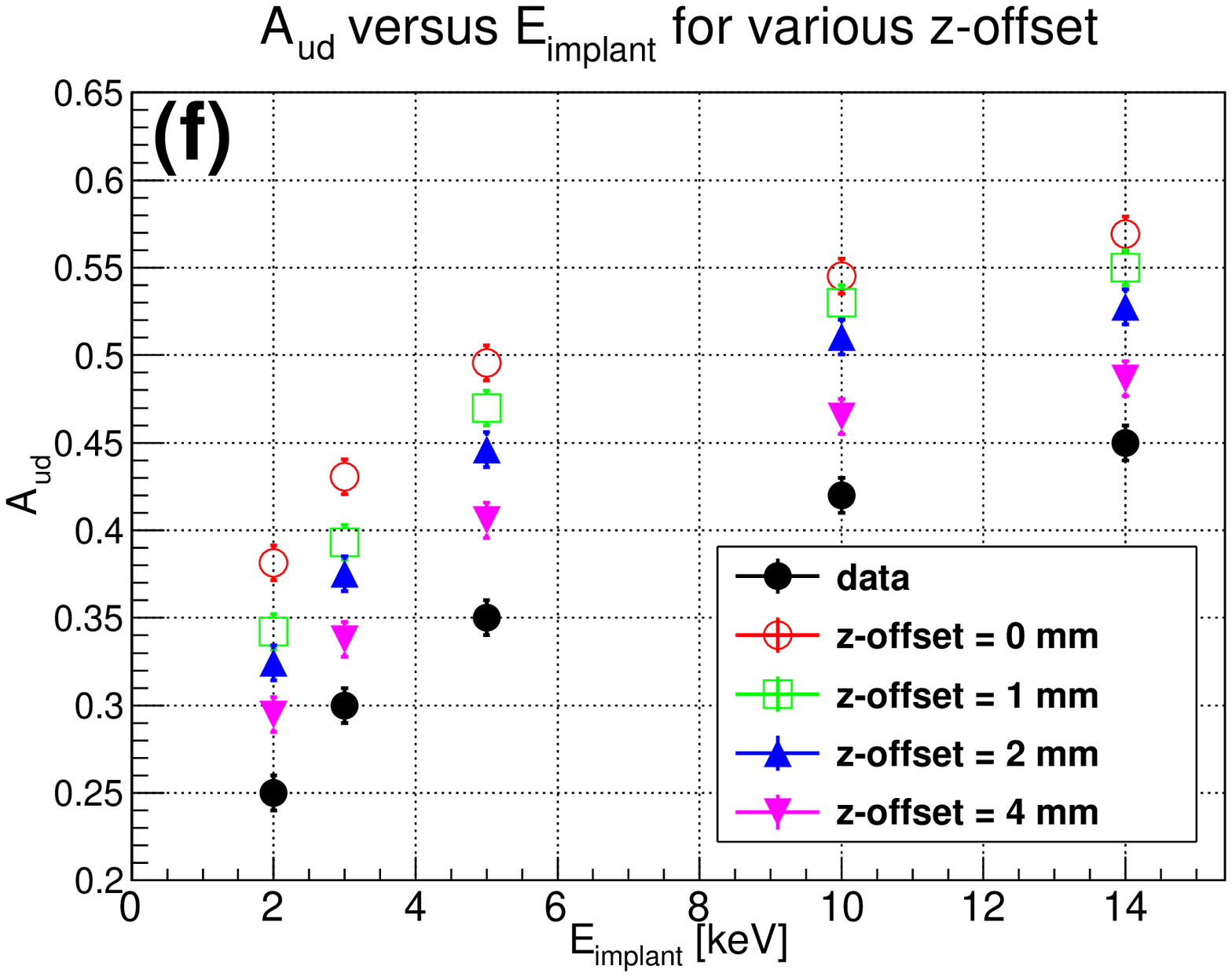}
\caption{Measured and simulated $A_{\rm{ud}}$ versus $\mu^{+}$
implantation energy. $\sigma_{xy}$ and $\sigma_{x'y'}$ are the
standard deviation of the muon phase space after the acceleration section
of the moderator,
$E^{SR}_{x}$ the electric field in the spin rotator and $V_{RA}$ the HV at the conical
lens. For each plot only one parameter is varied whereas the other
one are given in table~\protect\ref{tab1}.}
\label{fig:AUDvsAll}
\end{figure}

The dependence of $A_{\rm{ud}}$ on the various beam line parameters has
been investigated and summarized in figure~\ref{fig:AUDvsAll}.

\begin{figure}[htbp]
\centering
\includegraphics[width=0.95\textwidth]{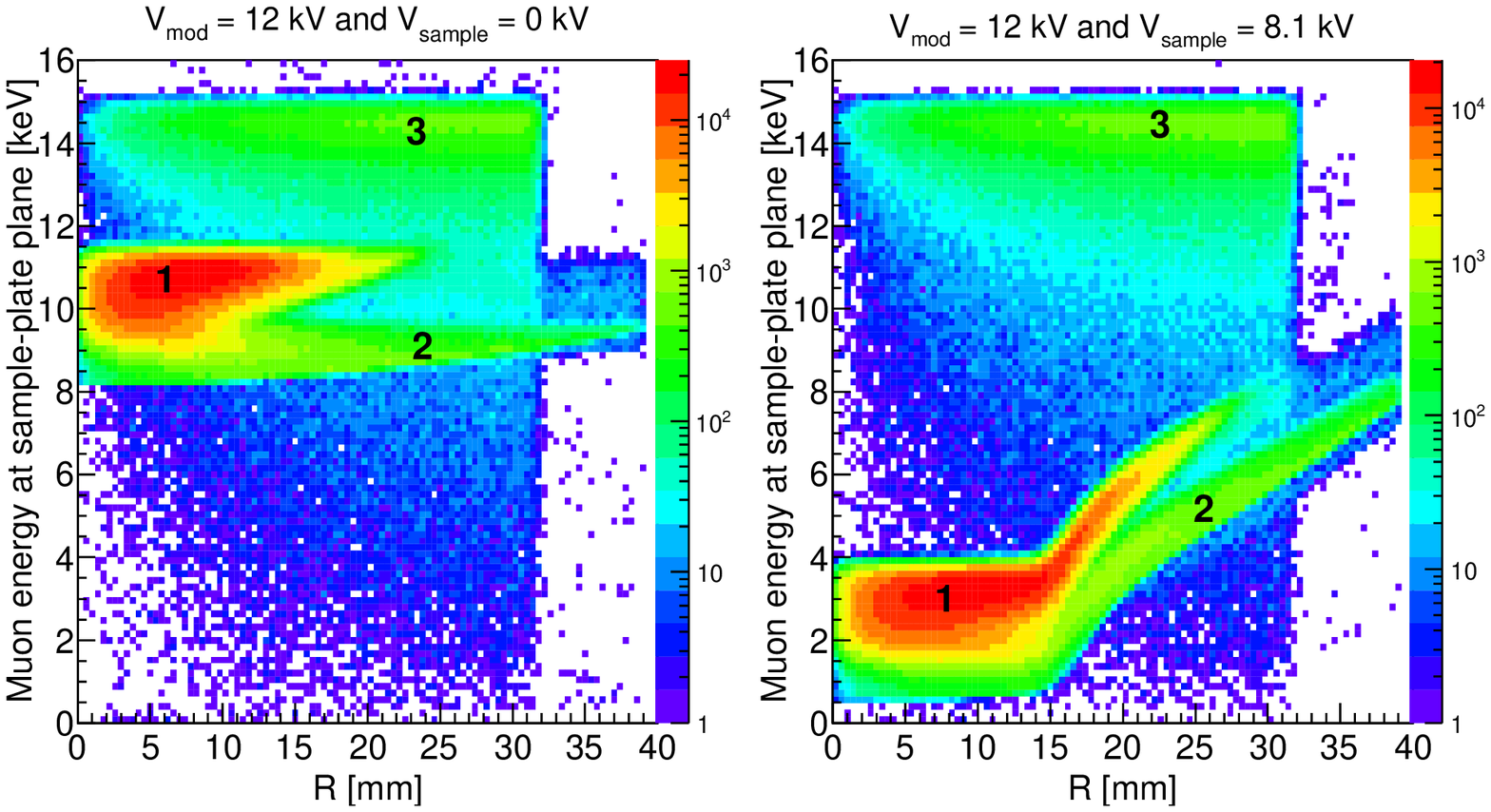}
\caption{Simulated energy of the muon beam at sample-plate plane as a function of distance $R$ from the beam axis
for 2 different muon implantation energies: (a) 11~keV and (b) 3~keV. Region 1 is mainly contributed by
muons with kinetic energy above 9.15~keV before reaching the RA. 
Region 2 is mainly given by muons with energy around 9.15~keV ($=V_{\rm{RA}}$) before reaching the RA.
Region 3 is originated from Mu, which are not decelerated by the electric fields in TD and are not affected by the RA and sample
plate electric potentials, and hence have higher energies than the $\mu^{+}$. Since Mu are not focused by the RA lens they have a distribution with larger transverse extension.}
\label{fig:ke_R}
\end{figure}

\begin{figure}[tbp]
\centering
\includegraphics[width=0.499\textwidth]{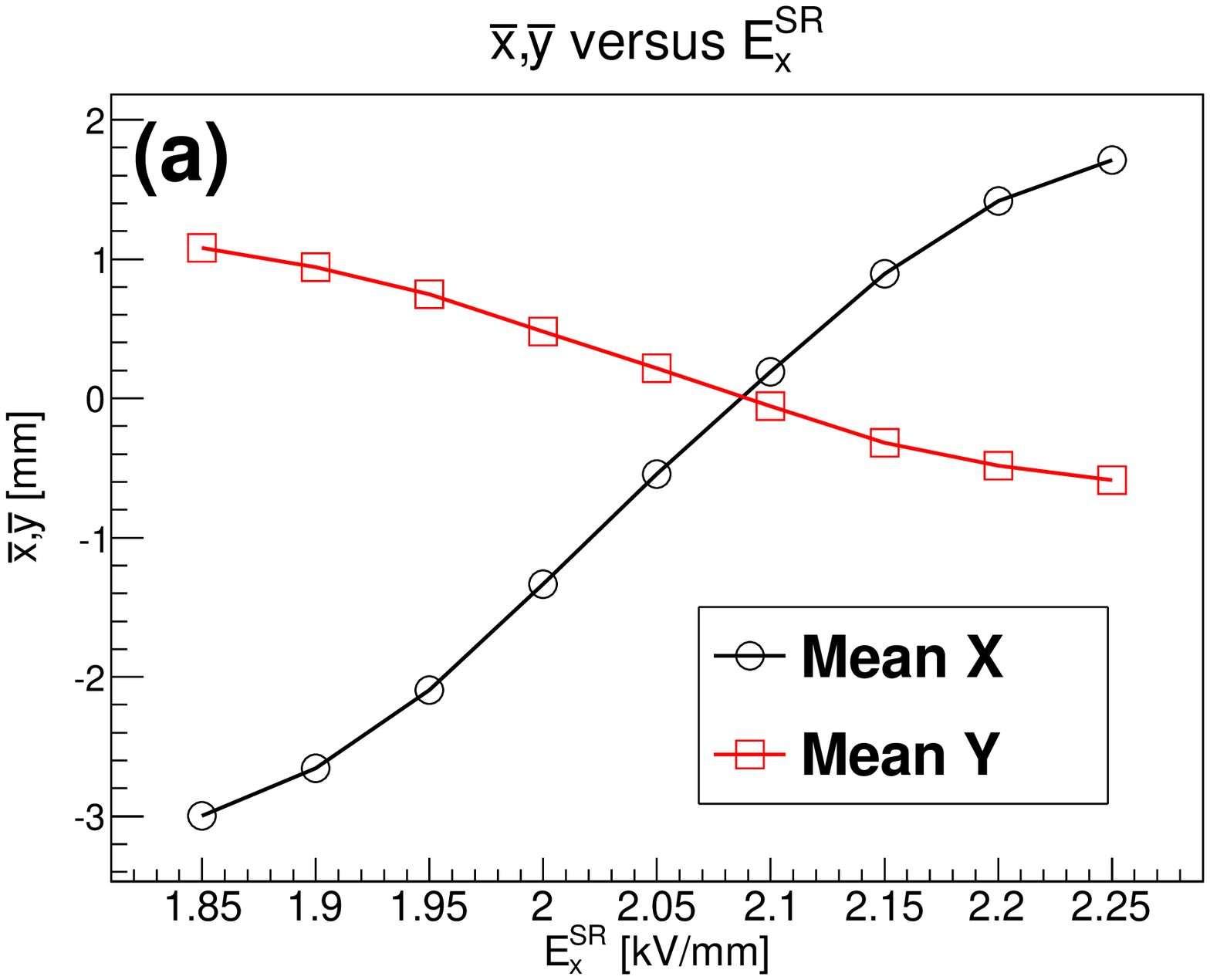}
\includegraphics[width=0.495\textwidth]{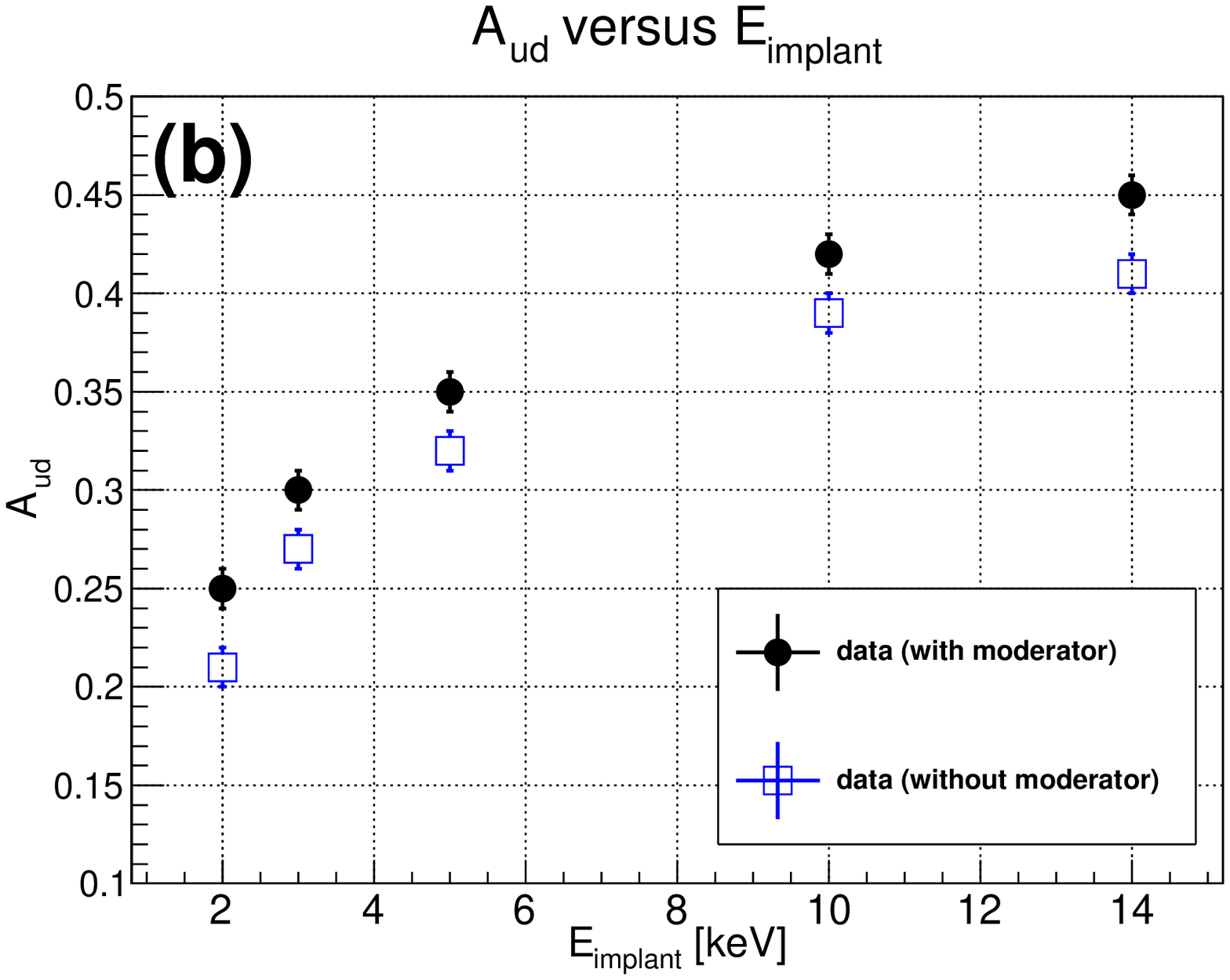}
\caption{(a) Simulated $\mu^{+}$ beam mean position
  ($\overline{x}$,$\overline{y}$) at the sample plate as a function of
  the electric field in the spin rotator $E^{SR}_{x}$. The
  $E^{SR}_{x}$ value in table~\protect\ref{tab1} is chosen such that the beam is centered on the
  original sample plate position (16~mm downstream of the nose sample
  plate). (b) Measured $A_{\rm{ud}}$ versus $\mu^{+}$
  implantation energy  with and without Ar-N$_{2}$ layer
  at the moderator. ``Unmoderated'' muons at the plate position have a
  larger beam size.}
\label{fig:AUDvsEimplantUnMod}
\end{figure}

\begin{itemize}
\item{ \bf (a) Muon implantation energy:}
  $A_{\rm{ud}}$ is increasing with decreasing implantation energy
  due to defocussing effects caused by the electric potential applied
  at the sample holder (see figure~\ref{fig:Efieldcomsol}(a)).
  The musrSim (new) represents the Geant4 simulation where
  the muon energy losses at the C-foil are parametrized using
  Landau distributions. From figure~\ref{fig:ke_R}, it can be seen that
  for a lower implantation energy, there is a considerable amount of $\mu^{+}$
  which do not hit the ``nose sample plate'' which has a radius of 15~mm. 
  Thus at lower implantation energy, there is a higher probability of decay positron
  being detected by the downstream detector, resulting in a reduced $A_{\rm{ud}}$ asymmetry.

\item {\bf (b+c) Initial phase space of the muon beam:} Initial phase
  space ($xx'$,$yy'$) after the acceleration section of the moderator is
  taken to be $\sigma_{x}=\sigma_{y}=7.5$~mm for the beam size and
  $\sigma_{x'}=\sigma_{y'}=2.0^\circ$ for the beam divergence from a recent
  simulation~\cite{Hofer1998}.
The initial polarization vector is chosen as
$\mathbf{P_{\mu}}=(0.9848, 0, 0.17365)$ since the $\mu^{+}$ spin is
rotated by 10$^\circ$ clockwise after traversing the electrostatic
separator of the $\mu$E4 beam line before the $\mu^{+}$ is focused on
moderator (see figure~\ref{fig:lem2012}).

\item {\bf (d) Electric and magnetic field of the spin rotator:}
The magnetic field of the spin rotator was fixed to $B^{SR}_{z}=
-62.3(-69.7)$~G for 12(15) keV transport energy to obtain the
experimental $\mu^{+}$ spin rotation of 20$^\circ$
counter-clockwise. The spin angle with respect to the $\mu$E4 beam
direction (corresponding to -$x$ direction in the simulation) is then
changed from +10$^\circ$ (the angle after the separator in the $\mu$E4
beam line) to -10$^\circ$. The electric field was varied from
$E^{SR}_{x}=1.85$~to~2.25~kV/mm and is chosen such that the beam is
centered on the original sample plate position (16~mm downstream of
the nose sample plate) as shown in
figure~\ref{fig:AUDvsEimplantUnMod}(a). 
From figure~\ref{fig:AUDvsAll}(d) it is visible that
that $A_{\rm{ud}}$ is maximal at around 2.09~kV/mm. This occurs
when the beam is centered on the sample plate (see figure~\ref{fig:AUDvsEimplantUnMod}(a)) and the downstream
detectors are shielded from the positron by the material of the ``nose sample
plate'' itself.

The design value of 2.68~kV/mm which was expected from simply
considering the relation $v=\frac{E}{B}$ has thus to be tuned to
2.09~kV/mm to center the beam on the axis.
This indicates that there are not well understood imperfections of the
electric and magnetic fields that cause the $\mu^{+}$ to deviate from
its original path.

\item {\bf (e) Electric potential of the conical lens (RA):} If the
  focusing power is not optimal the $A_{\rm{ud}}$ decreases since
  the beam becomes larger.

\item {\bf (f) $Z$-offset of the sample plate position:} It is obvious
  that there is a strong dependence of the $A_{\rm{ud}}$ on the
  z-position of the ``nose sample plate''. Moving the sample plate
  downstream, in z-direction, will decrease the $A_{\rm{ud}}$ as
  the downstream detector will be less shielded by the sample plate
  from positrons.

\end{itemize}

Summarizing, the new musrSim with Landau distributed losses give
asymmetries closer to the measured one.
However, the agreement is not yet satisfactory.
Some parameters could be slightly tuned around the design value to
decrease the deviation between measured and simulated
$A_{\rm{ud}}$ asymmetries.
The newly inserted spin rotator optimal settings do not correspond to
the design value manifesting that some uncontrolled beam distortion
and beam transmission may occur at this beam line element.
Contribution of ``unmoderated'' $\mu^{+}$ is shown in figure~\ref{fig:AUDvsEimplantUnMod}(b).
As they have a larger beam spot ($x_{RMS}$ and $y_{RMS}$) as shown in
table~\ref{beamspot} and hence a lower value of $A_{\rm{ud}}$.
The larger beam spot could be caused by chromatic aberration in the beam transport optics 
due to the wide energy distribution of the ``unmoderated'' $\mu^{+}$.
But this is not enough to explain the observed discrepancy between experimental and simulated
$A_{\rm{ud}}$ asymmetries as they accounted for only 10-15\% of the measured time spectra.
However, as shown in figure~\ref{fig:AUDvsAll}, by tweaking
different parameters one could get agreement.
We refrain at this stage to perform a multi-parameter tuning because
of the complexity and the correlation between the various parameter.
\begin{table}[htb]
\caption{Measured $\mu^{+}$ beam spot $x_{RMS}$ and
  $y_{RMS}$ for the ``unmoderated'' $\mu^{+}$ and ``moderated'' $\mu^{+}$ using the MCP2. It
  should be noted that due to the finite size of the active region of
  the MCP2 (42~mm in diameter) the actual $x_{RMS}$ and
  $y_{RMS}$ could be larger, especially for the ``unmoderated''
  $\mu^{+}$.}  \centering
\begin{tabular}{|c|c|c|c|c|c|} \hline
Energy (keV) & Spin rotator & Moderator & $x_{RMS}$ (mm) & $y_{RMS}$ (mm) & $\mu^{+}$ type \\ \hline
15 & No & No & 6.1 & 6.0 & unmoderated \\ \hline
15 & No & Yes & 5.3 & 4.9  & moderated\\ \hline
15 & Yes & No & 7.2 & 7.2  & unmoderated \\ \hline
15 & Yes & Yes & 6.2 & 6.0 & moderated \\ \hline
\end{tabular}
\label{beamspot}
\end{table}

The reason why there was a good agreement between the simulated and
measured $A_{\rm{ud}}$ prior to the LEM beam line upgrade, is
related with the smaller beam size at the sample plate which
imply a reduced defocussing effect at the ``nose sample plate''.
The insertion of the spin rotator has caused a degradation of the beam
quality and an unexpected increase of the beam size at the sample
position.
This increase could be even more substantial for the ``unmoderated'' muon
component.
The larger RMS values ($\approx$~1.2~mm) for ``unmoderated'' muons could be
sufficient to explain the observed smaller asymmetries in figure~\ref{fig:AUDvsAll}.

\section{Conclusions}

Simulations of the complete LEM beam line after the 2012 upgrade have
been presented.
TOF measurements have been used to calibrate the energy losses in the
start detector.
Excellent agreement between the measured TOF spectra and simulations
has been reached only by using Landau distributed energy straggling,
accounting for muonium production, and accounting for a
contribution of ``unmoderated'' muons with slightly larger kinetic
energy which are parasitically transported by the beam line.
The measured stopping power in the C-foil compares well with previous
determination.
This good agreement between measured and simulated TOF spectra implies
also the correctness of the assumed muonium yield in the C-foil which
has been implemented using velocity scaling of proton data.

Detailed analysis of the $\mu$SR time spectra require information of
the muon arrival times, muon implantation energy and depolarization
effects related with energy losses and Mu production in the C-foil.
These information can be determined now for all beam line settings and
muon implantation energies using the new Geant4 simulations.

The beam spot size at the sample position is also a very important
parameter when analyzing the $\mu$SR data.  
Normalizations and total measurable decay asymmetries depend on this parameter.
This is even more important for longitudinal $\mu$SR which is now
possible due to the insertion of the spin rotator.
Therefore the beam spot size at the sample position has been
investigated by means of the $A_{\rm{ud}}$ asymmetry using a
dedicated sample plate (``nose sample plate'') with increased
sensitivity to beam changes.
These studies have revealed a problem with the beam transport in the
LEM beam line related with the newly inserted spin rotator and the
parasitic transport of ``unmoderated'' muons which has called for 
detailed studies and hardware improvements.

\acknowledgments This work has been supported by the Swiss National
Science Foundation under the grant numbers 200020\_146902 and PZ00P2\_132059.
We also acknowledge the help of the PSI and ETH Zurich IPP workshops and support
groups. Special thanks to T. Shiroka, V. Vrankovi\'{c} and M. Horisberger.

\end{document}